\documentclass[12pt]{article}
\usepackage{amsmath}
\usepackage{amsfonts}
\usepackage{amssymb}
\usepackage{srcltx}

\newtheorem{theorem}{Theorem}
\newtheorem{proposition}[theorem]{Proposition}
\newtheorem{lemma}[theorem]{Lemma}
\newtheorem{corollary}[theorem]{Corollary}

\newtheorem{remark}[theorem]{Remark}
\newtheorem{example}[theorem]{Example}

\begin{document}
\noindent
{\Large {\bf The generalized Giambelli formula and polynomial KP and CKP tau-functions}}
\vskip 9 mm
\begin{minipage}[t]{70mm}
{\bf Victor Kac}\\
\\
Department of Mathematics,\\
Massachusetts Institute of Technology,\\
Cambridge, Massachusetts 02139, U.S.A\\
e-mail: kac@math.mit.edu\\
\end{minipage}\qquad
\begin{minipage}[t]{70mm} 
{\bf Johan van de Leur}\\
\\
Mathematical Institute,\\
Utrecht University,\\
P.O. Box 80010, 3508 TA Utrecht,\\
The Netherlands\\
e-mail: J.W.vandeLeur@uu.nl
\end{minipage}
\\
\
\\
\begin{abstract}
The first part of the paper is devoted to two descriptions of all polynomial 
tau-functions
of the KP hierarchy: by a generalized Jacobi-Trudy formula, and a 
generalized Giambelli formula.
We use the latter formula in the second part to obtain
 all polynomial tau-functions of the CKP hierarchy and its $n$-reductions.    In particular,  for $n=3$ we find all polynomial tau-functions of the Kaup-Kupershmidt hierarchy. 
\end{abstract}
\section{Introduction}
The concept of a tau-function of a hierarchy of soliton equations, developed by the Kyoto school in early 80's (see \cite{S},\cite{DJKM2},\cite{JM}) is very
useful for construction of solutions of these equations.

The geometric meaning of a tau-function is very simple: it is, up to a constant factor, a non-zero element of the orbit of a highest weight vector of  a representation of an infinite-dimensional group \cite{S},\cite{DJKM2},\cite{JM},\cite{KPeterson}.

The first and the most famous example is the KP hierarchy, constructed as follows. Let $C\ell$ be the associative algebra on generators $\psi^+_j$ and $\psi^-_j$, $j\in\frac12+\mathbb Z$,  subject to the relations 
\begin{equation}
\label{1.1}
[\psi^+(z),\psi^-(w)]_+=\delta(z-w),\quad [\psi^\pm(z),\psi^\pm(w)]_+=0,
\end{equation}
where 
\begin{equation}
\label{psi}
\psi^{\pm}(z)=\sum_{i\in\frac12`+\mathbb Z} \psi_i^\pm z^{-i-\frac12}
\end{equation}
are the generating series,  called the charged free fermionic fields,
and $\delta(z-w)=z^{-1}\sum_{n\in\mathbb Z}\left(\frac{z}{w}\right)^n$ is the formal delta function.
 Let $F$ be an irreducible representation of the algebra $C\ell$, which admits a non-zero vector $|0\rangle$,  such that
 \begin{equation}
\label{1.2}
\psi^{\pm}_{j} |0\rangle = 0, \ \text{for}\ j > 0 . 
\end{equation}
 
Let $GL_\infty$ be the group 
of   matrices  $ (g_{ij})_{i,j \in {\frac12+\mathbb Z}}$  with entries in $\mathbb C$,  which are 
 invertible and all, but a finite number of $g_{ij} -
\delta_{ij}$, are $0$.  
We obtain a representation $R$ of this group on $F$ by letting
\begin{equation}
\label{1.3}
R(I+aE_{ij})=1+a\psi^+_{-i}\psi^-_j,  \quad i,j\in \frac12+\mathbb Z, \quad a\in\mathbb C.
\end{equation}
Defining  the charge decomposition
\begin{equation}
\label{1.3a}
F = \bigoplus_{m \in {\mathbb Z}} F^{(m)}, \
\end{equation}
by letting 
\[
\text{charge}(|0\rangle ) = 0\ \text{and charge} (\psi^{\pm}_{j}) =
\pm 1,
\]
we see that each $F^{(m)}$ is an irreducible
  highest weight
 module over $GL_\infty$, and 
 \begin{equation}
 \label{1.4}
 |\pm m\rangle=\psi^\pm_{-\frac{2m-1}2}\cdots\psi^\pm_{-\frac32}\psi^\pm_{-\frac12}
 |0\rangle, \quad m\in\mathbb Z_{\ge 0},
 \end{equation}
 is a highest weight vector for $F^{(\pm m)}$.
 
 The KP hierarchy in the fermionic picture 
 is defined as the following equation:
 \begin{equation}
 \label{1.5}
{\rm Res}_{z}\,\psi^+(z)\tau\otimes\psi^-(z)\tau =0,\quad \tau\in F^{(0)},
\end{equation}
where ${\rm Res}_z\, \sum_if_i z^i=f_{-1}$.
It is easy to show \cite{KPeterson} that equation \eqref{1.5} holds for a non-zero 
$\tau\in F^{(0)}$ if and only if $\tau$ lies in the $R(GL_\infty)$-orbit of $|0\rangle$.

A remarkable fact is that equation \eqref{1.5} can be converted in a collection of PDE's, using bosonization of $F$.  For this one introduces the free  
  bosonic field
\begin{equation}
\label{alpha}
\alpha(z)=\sum_{n\in\mathbb Z}\alpha_nz^{-n-1}=:\psi^+(z)\psi^-(z):,
\end{equation}
where,  as usual,  $:\psi^+_i\psi^-_j:=\psi^+_i\psi^-_j$ if $i\le j $, and 
$=-\psi^-_j\psi^+_i$ if $i>j$.  Then the $\alpha_n$ 
satisfy the commutation relations of the
infinite Heisenberg Lie algebra
\begin{equation}
\label{Heis}
[\alpha_m,\alpha_n  ]=m\delta_{m,-n}, 
\end{equation}
and  since $\alpha_i|0\rangle=0$ for $i\ge 0$,
  there exists an isomorphism
$$\sigma: F\xrightarrow{\sim} \mathbb C[q,q^{-1}, t_1,t_2,\ldots],$$
called the bosonization of $F$,  which
is uniquely determined
by the following properties
\begin{equation}
\label{sigma-alpha}
\sigma (|m\rangle)=q^m,\
\sigma\alpha_0\sigma^{-1}=q\frac{\partial}{\partial q},
\ 
\sigma\alpha_{-i}\sigma^{-1}=it_i \ \mbox{and }
\ 
\sigma\alpha_i\sigma^{-1}=\frac{\partial}{\partial t_i} \ \mbox{for }i>0.
\end{equation}
Furthermore, since $[\alpha_k,\psi^\pm (z)]=\pm z^k\psi^\pm(z)$,
one can identify, under the isomorphism 
$\sigma$,
the charged free fermions with the vertex operator
\begin{equation}
\label{charged-vertex}
\sigma \psi^{\pm }(z)\sigma^{-1}
=
q^{\pm 1}z^{\pm q\frac{\partial}{\partial q}}
\exp\left(\pm \sum_{i=1}^\infty t_iz^i\right)
\exp\left(\mp \sum_{i=1}^\infty \frac{\partial}{\partial  t_i}\frac{z^i}{i}\right).
\end{equation}
 Then equation \eqref{1.5} gets converted to the KP hierarchy of bilinear PDE's on $\tau(t)\in\mathbb C[t_1,t_2,\ldots]$:
\begin{equation}
\label{bosonic-0KP}
{\rm Res}_{z}\
\exp\left( \sum_{i=1}^\infty (t_i-{t'}_i)z^i\right)
\exp\left( \sum_{i=1}^\infty \left(\frac{\partial}{\partial  {t'}_i}-
\frac{\partial}{\partial  t_i}\right)
\frac{z^{-i}}{i}\right)\tau(t)\tau(t')=0.
\end{equation}
Here and further $t'$ denotes another copy of $t=(t_1,t_2,\ldots)$.

Next, equation \eqref{bosonic-0KP} can be rewritten in terms of Lax type equations via 
the dressing operators $P(t,\partial)$, where $\partial =\frac{\partial}{\partial t_1}$ \cite{S}.
This is a monic pseudodifferential operator, whose symbol is 
\[
P(t,z)=\frac{\exp (- \sum_{i\ge 1} \frac {z^{-i}}i   \frac{\partial}{\partial  t_i} )\tau(t)
}
{\tau (t)}.
\]
The  associated to $\tau(t)$ Lax operator $L(t, \partial)$ is defined as the pseudodifferential operator 
\begin{equation}
\label{1.11}
L(t, \partial)=P(t, \partial)\circ  \partial\circ P(t, \partial)^{-1}.
\end{equation}
Then equation \eqref{bosonic-0KP} on the tau-function $\tau(t)$ is equaivalent to the  following hierarchy 
of Lax-Sato evolution PDE's on 
$L(t,\partial)=\partial +\sum_{j>0}u_j(t)\partial^{-j}$:
\begin{equation}
\label{Laxeq}
\frac{\partial L (t,\partial)}{\partial t_k}=[(L (t,\partial)^k)_+, L (t,\partial)],\quad k=1,2,3,\ldots ,
\end{equation}
where  the subscript$\ _+$, as usual,  denotes the differential part of $L (t,\partial)^k$.

A famous result of Sato \cite{S} is that  all Schur polynomials $s_\lambda(t)$
are tau-functions
of the KP hierarchy.  Recall that the Schur polynolial $s_\lambda(t)$, associated to a partition 
$\lambda=(\lambda_1\ge \cdots\ge \lambda_\ell >0)$ is defined by the Jacobi-Trudi formula
(see e.g.   \cite{macdonald}, Section I.3):
\begin{equation}
\label{Jacobi-Trudi}
s_\lambda(t)=\det \left(s_{\lambda_i+j-i} (t)\right)_{1\le i,j\le \ell},
\end{equation}
where the elementary Schur polynomials $s_j(t)$
are defined by the generating series
\begin{equation}
\label{el-schur}
\sum_{j=0}^\infty s_j(t)z^j=
\exp \sum_{i=1}^\infty t_iz^i.
\end{equation}

In our paper \cite{KvdLmodKP} we proved that all polynomial  tau-functions of the KP hierarchy    are,  up to a   constant factor,  of the form
\begin{equation}
\label{tau-lambdaintro}
\tau_{\lambda,c}(t )=\det \left(s_{\lambda_i+j-i} (t_1+c_{1i},t_2+c_{2i},t_3+c_{3i}, \ldots)
\right)_{1\le i,j\le \ell},
\end{equation}
where $\lambda=(\lambda_1, \lambda_2,\ldots ,\lambda_\ell) $,    and  $c=(c_{ij})$ is a 
$
 (\lambda_1+\ell-1)\times\ell$  matrix over $\mathbb C$.
We call equation \eqref{tau-lambdaintro}
the generalized Jacobi-Trudi formula for polynomial KP tau-functions.

It is well known that,  using the Frobenius notation 
$\lambda=(a_1,a_2,\ldots ,a_k|b_1,b_2, \ldots, b_k)$
for the partition $\lambda$, one can write the Schur polynomial $s_\lambda(t) $
in the Giambelli form  (see e.g.   \cite{macdonald}, Section I.3):
\begin{equation}
\label{Giam}
s_\lambda(t)=\det \left(\chi_{(a_i|b_j)} (t;t)\right)_{1\le i,j\le k},
\end{equation}
where 
\begin{equation}
\label{chi}
\chi_{(a |b)}(t;t'):=(-1)^b\sum_{n=0}^{b}
 s_{n+a+1}(t)
  s_{b-n}(-t'),\qquad a,b\in\mathbb Z_{\ge 0}.
\end{equation}
The first new result of the paper is Theorem 
\ref {T-Giambelli} in Section \ref{S3}, describing all polynomial KP tau-functions  
by the generalized Giambelli formula (see also \cite{HL} and \cite{HO}):
\begin{equation}
\label{tau-lambda2intro} 
\tau_{\lambda;c,d}(t)=\det \left(\chi_{(a_i|b_j)} (t_1+c_{1i},t_2+c_{2i},  \ldots;t_1+d_{1j},t_2+d_{2j}, \ldots)
\right)_{1\le i,j\le k},
\end{equation}
 where $c=(c_{ij})$ is a $(a_1+b_1+1)\times k$ matrix over $\mathbb C$ and $d=(d_{ij})$ is a $b_1\times k$ matrix over $\mathbb C$.

In Section \ref{S4}, using the Jacobi-Trudi formalism, as in \cite{KRvdL}, we construct more general KP tau-functions, and find a Jacobi-Trudi type formula for the 
wave function
$w^+(t,z)=P(t,\partial) \exp \sum_{i=1}^\infty t_iz^i$.
 We also find analogous formulas in the framework of the Giambelli formalism.
 
 Our Theorem 
  \ref{prop5} describes,  in particular, the polynomial tau-functions 
 for the CKP hierarchy.
 Recall that the CKP hierarchy in the Lax-Sato form is the following hierarchy of evolution equations on the skew-adjoint pseudodifferential operator
 $L(t_o,\partial)=\partial +\sum_{j>0}a_j(t_o)\partial^{-j}$,
 where $t_o=(t_1,0,t_3,0,t_5,\ldots)$ \cite{DJKMVI}:
  \begin{equation}
 \label{1.18}
 \frac{\partial L (t_o,\partial)}{\partial t_k}=[(L (t_o,\partial)^k)_+, L (t_o,\partial)],\quad k=1,3,5,\ldots .
 \end{equation}

 There are at least two ways to construct the corresponding tau-function.
 One is to use the construction of the metaplectic representation of the infinite symplectic group $SP_\infty$ via symplectic bosons, as in \cite{LOS} and \cite{KLlang}.
 However, in this paper we use another way, via the reduction of 
 the representation  of $Gl_\infty$ in $F^{(0)}$  to $SP_\infty$ \cite{JM} (see also  \cite{KZ},  \cite{HHH})  
 
 Let $\mathbb C^\infty=\bigoplus_{i\in\frac12+\mathbb C} \mathbb C e_i$,  so that $GL_\infty$ is the group of automorphism of this vector space,   leaving all but a finite number of the $e_j$ fixed.  Define a skew-symmetric bilinear form $(\cdot, \cdot )_C $
 on $\mathbb C^\infty$ by 
 \begin{equation}
 \label{1.19}
 (e_i,e_j)_C=(-1)^{i+\frac12}\delta_{i,-j}.
 \end{equation}
 Then  $SP_\infty$ is the subgroup of $GL_\infty$,  leaving this bilinear form invariant.
 
 Define the automorphism  $\iota_C$ of the algebra $C\ell$ by \cite{JM}
\begin{equation}
\label{1.20}
\iota_C(\psi_j^\pm)=(-1)^{j \pm \frac12}\psi^\mp_j.
\end{equation}
 This automorphism induces an automorphism of the vector space $F$, which we again denote by $\iota_C$, by letting  $ \iota_C(|0\rangle)=|0\rangle$.
 The subspace $F^{(0)}$ of $F$ is $\iota_C$-invariant,  and we denote by
 $F^{(0)}_C$ the fixed point set  of $\iota_C$ in $F^{(0)}$.  An element $\tau$ in the orbit $R(SP_\infty)|0\rangle$ then satisfies 
 the following equation 
 (cf.  \eqref{1.5})
 \begin{equation}
\label{1.21}
{\rm Res}_{z}\, \psi^+(z)\tau\otimes \psi^+(-z)\tau =0,
\end{equation}
 which is called the CKP hierarchy in the fermionic picture.
 After bosonization equation \eqref{1.21} becomes (cf. \eqref{bosonic-0KP}):
\begin{equation}
\label{bosonic-CKPintro}
{\rm Res}_{z=0}\, 
\exp\left( \sum_{i=1}^\infty (t_i+(-1)^i {t'}_i)z^i\right)
\exp\left(- \sum_{i=1}^\infty \left(
\frac{\partial}{\partial  t_i}+(-1)^i\frac{\partial}{\partial  {t'}_i}\right)
\frac{z^{-i}}{i}\right)\tau (t)\tau (t')=0.
\end{equation}
 A non-zero element $\tau(t)\in \mathbb C[t_1,t_2, \ldots ]$,
 satisfying 
 \eqref{bosonic-CKPintro}, is called  a tau-function of the CKP hierarchy,   if it satisfies 
 \begin{equation}
\label{tauCintro}
\tau(t_1,t_2,t_3,t_4,\ldots)=\iota_C(\tau(t_1,t_2,t_3,t_4,\ldots))=\tau(t_1,-t_2,t_3,-t_4,\ldots),
\end{equation}
since,  under the bosonization we  have 
$$\sigma: F^{(0)}_C\xrightarrow{\sim} B_C:=\{ f\in
\mathbb 
C[ t_1,t_2,\ldots]|\, \iota_C(f)=f\}
.$$
In order to obtain a skew-adjoint Lax operator $L(t_o,\partial)$
from  $L(t,\partial)$, satisfying \eqref{1.11} \eqref{Laxeq},
we substitute $t_{2j}=0$,  $j=1,2,3,\ldots$,  in the CKP tau-function $\tau(t)$.

Our Theorem \ref{prop5}   describes, in particular,   all polynomial tau-functions of the  
CKP hierarchy   in the Giambelli form.
Namely, they correspond to self-conjugate partitions, which in the Frobenius notation are $\lambda=(a_1,a_2,\ldots,a_k|a_1,a_2,\ldots,a_k)$, and are of the form \eqref{tau-lambda2intro},  where $d=\iota_C(\overline c)$,   
$\overline c$ consists of the first $b_1$ rows of $c$,  and
$\iota_C$ stands for changing the sign of even numbered rows of the matrix $\overline c$; in addition, the matrix $c$ must satisfy
the constraint \eqref{restrict2} in Section \ref{S5} (which holds for $c=0$).

In Section \ref{S6}
 we prove Theorem \ref{T12} on polynomial tau-functions for the $n$-reduced CKP hierarchy, using the results on the polynomial   tau-functions of the KP hierarchy  and the  $n$-reduced KP hierarchies.
The $2$-reduced CKP hierarchy \eqref{Laxeqred} is just the KdV hierarchy.
The $3$-reduced CKP hierarchy is called the Kaup-Kupershmidt hierarchy. 
It is a hierarchy of evolution PDE's on  the function 
\begin{equation}
\label{25a}
u(t_o) =3\frac{\partial^2\log \tau(t_o)}{\partial t_1^2},
\end{equation}
where $t_o=(t_1,t_3,t_5,\ldots)$,
written as Lax equations on the differential operator
\begin{equation}
\label{87c}
{\cal L}(t_o,\partial)=\partial^3+u\partial+\frac 12 u_x,\ \mbox{
where }
 x=t_1,
 \end{equation}
 namely
\begin{equation}
\label{87a} 
\frac{
\partial {\cal L}(t_o,\partial)}{\partial t_k}=
[({\cal L}(t_o, \partial)^{\frac{k}3})_+
, {\cal L}(t_o, \partial )]
,\quad k=1,3,5,\ldots .
\end{equation}
If $k$ is  a multiple of 3, then $\frac{\partial u}{\partial t_k}=0$, hence the first non-trivial equation is \eqref{87a} for $k=5$,  and it is the Kaup-Kupershmidt equation \eqref{87b} in section \ref{S6}.

In the conclusion of  the paper we compare our results on polynomial 
tau-functions of the CKP hierarchy with that of the BKP    hierarchy,  found in 
\cite{KvdLB2}, \cite{KRvdL}, \cite{L}.

\section{The  KP hierarchy}
First, we briefly recall the basics of the theory of the KP hierarchy,  see \cite{DJKM2}, \cite{JM}, \cite{KvdLmodKP}, \cite{KPeterson}.
Consider the infinite  matrix group 
$GL_{\infty}$    
(resp.  its Lie algebra $gl_{\infty}$)
 consisting of all infinite  matrices  $G= (g_{ij})_{i,j \in {\frac12+\mathbb Z}}$  with entries in $\mathbb C$, which are 
 invertible and all but a finite number of $g_{ij} -
\delta_{ij}$ are $0$  (resp.  consisting of all matrices  $g= (g_{ij})_{i,j \in {\frac12+\mathbb Z}}$ for which are  all but a finite number of $g_{ij}$ are $0$).  Both 
act   on the vector space 
${\mathbb C}^{\infty} = \bigoplus_{j \in {\frac12+\mathbb Z}} {\mathbb
C} e_{j}$  via the usual formula
$E_{ij} (e_{k}) = \delta_{jk} e_{i}$.  

The semi-infinite wedge representation \cite{KPeterson}, \cite {KvdLmodKP} 
 $F =
\Lambda^{\frac{1}{2}\infty} {\mathbb C}^{\infty}$ is the vector space
with a basis consisting of all semi-infinite monomials of the form
$e_{i_{0}} \wedge e_{i_{1}} \wedge e_{i_{2}} \ldots$, where $i_{0} >
i_{1} > i_{2} > \ldots$ and $i_{\ell +1} = i_{\ell} -1$ for $\ell >>
0$.  One defines the  representation $R$ of $GL_{\infty}$ (resp.  $r$
of $gl_{\infty}$) on $F$ by
$$
\begin{aligned}
R(G) (e_{i_{0}} \wedge e_{i_{1}} \wedge e_{i_{2}} \wedge \cdots) &= G
e_{i_{0}} \wedge G e_{i_{1}} \wedge Ge_{i_{2}} \wedge \cdots  ,\quad &G\in GL_\infty,\\
r(g) (e_{i_{0}} \wedge e_{i_{1}} \wedge e_{i_{2}} \wedge \cdots)&=
\sum_{j=0}^\infty  e_{i_{0}} \wedge\cdots \wedge e_{i_{j-1}}\wedge
ge_{i_j}\wedge e_{i_{j+1}}\wedge\cdots \, ,\quad\quad &g\in gl_\infty,
\end{aligned}
$$
assuming the usual rules of the product $\wedge$.

The representation $r$ of the Lie algebra
$g\ell_{\infty}$ can be given 
in  terms of a Clifford algebra as follows.       Define the wedging and contracting
operators 
$\psi^{+}_{j}$ and $\psi^{-}_{j}$  $(j \in {\frac{1}{2}+\mathbb Z} 
)$ on $F$ by
\begin{equation}\label{111}\begin{aligned}
&\psi^{+}_{j} (e_{i_{0}} \wedge e_{i_{1}} \wedge \cdots ) =  
e_{-j }\wedge e_{i_{0}} \wedge e_{i_{1}} \cdots, \\
&\
\psi^{-}_{j} (e_{i_{0}} \wedge e_{i_{1}} \wedge \cdots ) = \begin{cases} 0
&\text{if}\ j  \neq i_{s}\ \text{for all}\ s \\
(-1)^{s} e_{i_{0}} \wedge e_{i_{1}} \wedge \cdots \wedge
e_{i_{s-1}} \wedge e_{i_{s+1}} \wedge \cdots &\text{if}\ j = i_{s}.
\end{cases}
\end{aligned}
\end{equation}
Then $r(E_{ij})=\psi^+_{-i} \psi^-_{j }$.
These operators satisfy the relations ( cf.\eqref{1.1})
$(i,j \in {\frac{1}{2}+\mathbb Z}, \lambda ,\mu = +,-)$:
\begin{equation}
\label{rel1}
\psi^{\lambda}_{i} \psi^{\mu}_{j} + \psi^{\mu}_{j}
\psi^{\lambda}_{i} = \delta_{\lambda ,-\mu} \delta_{i,-j}, 
\end{equation}
hence they generate a Clifford algebra, which we denote by ${\cal C}\ell$.
Introduce the following elements of $F$ $(m \in {\mathbb Z})$:
$$|m\rangle = e_{m-\frac12 } \wedge e_{m-\frac32 } \wedge
e_{m-\frac52 } \wedge \cdots .$$
It is clear that $F$ is an irreducible ${\cal C}\ell$-module such that the relations \ref{1.2}
hold.

It will be convenient to define also the oposite spin module with vacuum vector $\langle 0 |$, where
\[
\langle 0 |\psi^\pm_j=0,\ \text{for}\ j <0,
\]
and for $m>0$ one defines
\[
\langle \pm m|=\langle 0|\psi^\mp_{\frac12}\psi^\mp_{\frac32}\cdots\psi^\mp_{m-\frac12}.
\]
The vacuum expectation value is defined on $C\ell$ as $\langle a\rangle =\langle 0| a| 0\rangle$ and $\langle 0| 1| 0\rangle=1$.
Recall  the {\it charge decomposition} (\ref{1.3a},
the space $F^{(m)}$ is an irreducible highest weight
$g\ell_{\infty}$-module, where 
$|m\rangle$ is its highest weight vector, i.e.
$$
r(E_{ij})|m\rangle = 0 \ \text{for}\ i < j,\quad
r(E_{ii})|m\rangle = 0\  (\text{resp.}\ = |m\rangle ) \ \text{if}\ i > m\
(\text{resp. if}\ i < m).
$$

Let $S$ be the following operator on $F\otimes F$
\[
S=\sum_{i\in\frac{1}{2}+\mathbb{Z}
} \psi_i^+ \otimes \psi_{-i}^-
\]
and let 
${\cal O}_m
= R(GL_{\infty})|m\rangle \subset F^{(m)}$  be the $GL_{\infty}$-orbit
of the highest weight vector $|m\rangle$, then the following simple result  holds
\begin{theorem}
{\rm  \cite{KPeterson}}
Let $m$ be an integer and  let $0\ne f_m\in F^{(m)}$ .
Then  $f_m\in {\cal O}_m$ if and only if  
\begin{equation}
\label{mKP}
S(f_m\otimes f_m)=0.
\end{equation}
\end{theorem}

Equation (\ref{mKP}) is called the  KP hierarchy in the fermionic picture.  
 
To each $f_m= R(G ) |m\rangle\in {\cal O}_m$ one associates  a point in the Sato infinite Grassmannian 
 which is  the linear span of
 $\{ Ge_i|\, i<m\}\subset \mathbb C^\infty$.
 Another way to describe this subspace is as a subspace of $\Psi^+$,  where $\Psi^\pm=\bigoplus_{i\in \frac12+\mathbb Z}\mathbb C\psi^\pm_i$,  defined as the annihilation space ${\rm Ann}_+f_m$,  where 
 \[
{\rm Ann}_\pm f_m=\{v^\pm\in\Psi^\pm| v^\pm f_m=0\}.
 \]
 The connection between the two subspaces ${\rm Ann}_+f_m$ and ${\rm Ann}_-f_m$  is as follows (cf. \eqref{111}): If  $Ge_j=\sum_{i\in\frac12+\mathbb Z}G_{ij}e_i$,  then $G\psi^+_{-j}=\sum_{i\in\frac12+\mathbb Z}G_{ij}\psi^+_{-i}$,  and  ${\rm Ann_+f_m}$ is the linear span of $\{G\psi^+_i|\,  i<m\}$.
 We find ${\rm Ann}_- f_m$ as follows (see e.g. \cite{KvdLB}, Lemma 2.4):  
 ${\rm Ann}_- f_m$  is the linear span of        
 $\{G\psi^-_i|\,  i<-m\}$, and letting 
   $G\psi_{-k}^-=
 \sum_{i\in\frac12+\mathbb Z}H_{-i,k}\psi^-_i$,  since $(G\psi_{-j}^+,G\psi^-_k)=\delta_{jk}$,  we find that the matrix $(H_{ij})$ is the inverse transpose of $G$.

 Note that ${\rm Ann}  \, f_m={\rm Ann}_+ f_m\oplus {\rm Ann}_- f_m$ is a maximal isotropic subspace of 
 $\Psi=\Psi^+\oplus\Psi^-$, with respect to the symmetric bilinear form $(\,,\,)$, which defines the Clifford algebra $C\ell=C\ell(\Psi)$,
 \begin{equation}
 \label{bila}
  (\psi^+_i,\psi^-_j)=\delta_{i,-j}, \quad (\psi^\pm_i,\psi^\pm_j)=0.
  \end{equation} 
  \begin{proposition}
  {\rm \cite{KvdLB}}
  Let $f_m\in {\cal O}_m,$ and $v^\pm\in \Psi^\pm$,  such that $v^\pm f_m\ne 0$,  then $v^\pm f_m\in {\cal O}_{m\pm 1}$.
  \end{proposition}

We can extend the above description  to the Lie algebra $a_\infty$, which is the central extension  by a central element $K$ of the Lie algebra of infinite matrices  $(g_{ij})$ such that $g_{ij}=0$ if $|i-j|>>0$.   
The Lie bracket is given by $[g+\lambda K, h+\mu K] =gh-hg+C(g,h)K$,
where $C$ is the 2-cocycle 
given by
\[
\begin{aligned}
C(E_{ij},E_{ji})=1=-C(E_{ji},E_{ij})\ \mbox{if }i<0<j, \ \mbox{and }
C(E_{ij},E_{kl})=0\ \mbox{otherwise}.
\end{aligned}
\]
We obtain a representation $\hat r$ of $a_\infty$ on $F^{(m)}$, by the formula
\[
\hat r(E_{ij})=:\psi^+_{-i} \psi^-_{j }:,\quad \hat r (K)=1.
\]

Recall the free bosonic field $\alpha(z)=\sum_{n\in\mathbb Z}\alpha_n z^{-n-1}$, defined by
  \eqref{alpha}.  Then the operators $\alpha_n$ lie in $\hat r(a_\infty)$,  and satisfy the commutation relations 
  \eqref{Heis} of the infinite Heisenberg Lie algebra. Using this, one constructs the isomorphism  (bosonization)
$\sigma: F\xrightarrow{\sim} \mathbb C[q,q^{-1}, t_1,t_2,\ldots]$, uniquely defined by the properties \eqref{sigma-alpha}.
Recall that one has 
  \cite{DJKM2},\cite{JM}, \cite{KPeterson}
\begin{equation}
\label{charged-vertex2}
\sigma \psi^{\pm }(z)\sigma^{-1}
=q^{\pm 1}z^{\pm q\frac{\partial}{\partial q}}
\exp\left(\pm \sum_{i=1}^\infty t_iz^i\right)
\exp\left(\mp \sum_{i=1}^\infty \frac{\partial}{\partial  t_i}\frac{z^i}{i}\right).
\end{equation}
 
Let $f_m$   satisfy the KP hierarchy  in the fermionic picture \eqref{mKP},  and let $\tau_m=\sigma(f_m)$, then $\tau_m $  satisfies the following equation,  called the KP hierarchy of bilinear equations on $\tau_m(t)\in\mathbb C[t_1,t_2,\ldots]$:
\begin{equation}
\label{bosonic-mKP}
{\rm Res}_{z=0}\, 
\exp\left( \sum_{i=1}^\infty (t_i-{t'}_i)z^i\right)
\exp\left( \sum_{i=1}^\infty \left(\frac{\partial}{\partial  {t'}_i}-
\frac{\partial}{\partial  t_i}\right)
\frac{z^{-i}}{i}\right)\tau_m(t)\tau_m(t')=0.
\end{equation}
A solution $\tau_m(t)$ of \eqref{bosonic-mKP}
 is called a tau-function of the KP hierarchy.
A beautiful formula for the tau-function, corresponding to the point $R(G)|m\rangle$, where $G\in GL_\infty$, was given 
in \cite{DJKM2}:
\begin{equation}
\label{exptau}
\tau_m(t)=\langle m| (\exp  H(t))G| m\rangle ,
\end{equation}
where $H(t)=\sum_{i=1}^\infty t_i\alpha_i$.

\begin{remark}
\label{remarktau}
The totality of  tau-functions is independent of $m$.   Namely   if $G\in GL_\infty$ and $\Lambda=\sum_{i\in\frac12+\mathbb Z}E_{i,i+1}$, 
then $\Lambda^{-m}G\Lambda^m\in GL_\infty$  for any $m\in\mathbb Z$,   and
\[
\tau_m(t)=\langle m| \exp\left( H(t)\right)G| m\rangle = \langle 0| \exp\left( H(t)\right)  \Lambda^{-m}G\Lambda^m| 0\rangle.
\]
\end{remark}

Let $m=0$,  and $\tau(t)=\tau_0(t)$.  Define the wave  function $w^+(t,z)$ and the adjoint wave function $w^-(t,z)$ by
\begin{equation}
\label{wave1}
w^\pm(t,z)= \frac{\langle \pm 1|(\exp H(t))\psi^\pm (z)G|0\rangle}
  {\langle 0|(\exp H(t)) G|0\rangle}.
\end{equation}
Then 
\begin{equation}
\label{defwave}
w^\pm(t,z)= \frac{q^{\mp 1} (\sigma\psi^\pm (z)\sigma^{-1})\tau(t)}{\tau(t)}=
\frac{\tau(t\mp[z^{-1}])}{\tau(t)}\exp\left(\pm\sum_{n=1}^\infty t_nz^n\right),
\end{equation}
where $[z^{-1}]=(\frac{z^{-1}}1,\frac{z^{-2}}2,\frac{z^{-3}}3,\ldots)$,  and 
equation  \eqref{bosonic-mKP} is equivalent to the equation \cite{DJKM2}
\begin{equation}
\label{bil-wave}
{\rm Res}_z\, w^+(t,z)w^-(s,z)=0.
\end{equation}
One  can write this in terms of  monic pseudo-differential operators $P^\pm (t,\partial)$ in $\partial=\frac{\partial}{\partial t_1}$. Namely write \cite{DJKM2}
\begin{equation}
\label{def P}
\begin{aligned}
w^\pm (t,z)&=P^\pm (t,\pm z)\exp\left(\pm\sum_{n=1}^\infty t_nz^n\right)\\
&=P^\pm (t,  \partial)\exp\left(\pm\sum_{n=1}^\infty t_nz^n\right),
\end{aligned}
\end{equation}
then it is straightforward, see e.g.  \cite{KvdLmodKP}, Sections 3 and 4,  to prove that ($k=1,2,3,\ldots$):
\[
P^-(t,\partial)^*=P^+(t,\partial)^{-1}\quad\mbox{ and}\quad
\frac{\partial P^+(t,\partial)}{\partial t_k}=- (P^+(t,\partial)
\circ \partial^k\circ P^+(t,\partial)^{-1})_-\circ P^+(t,\partial).
\]
Using these equations, one  finds   that the KP Lax-Sato pseudodifferential operator $L(t,\partial)=P^+(t,\partial)\circ \partial \circ P^+(t,\partial)^{-1}$
satisfies the Lax-Sato evolution equations \eqref{Laxeq}.

\section{Polynomial KP tau-functions and the generalized Jacobi-Trudi and Giambelli formulas}
\label{S3}
Let $Par_\ell$ denote the set of partitions consisting of $\ell\in\mathbb Z_{\ge 0}$ non-zero parts, i.e.  sequences of integers 
$\lambda=(\lambda_1\ge \lambda_2\ge \cdots\ge \lambda_\ell>0)$, and let $\lambda=(a_1,\ldots, a_k|b_1,\ldots,b_k)$ be the Frobenius notation for $\lambda$ (see e.g.   \cite{macdonald}, Section I.1).  Let  $Par=\cup_{\ell=0}^\infty Par_\ell$.

Recall that the Schur polynomials  are given by the Jacobi-Trudi formula \eqref{Jacobi-Trudi}.
They are also given by the Giambelli formula \eqref{Giam} in the Frobenius notation of $\lambda$.
One of the main results of \cite{KvdLmodKP}, is Theorem 16, which describes all  polynomial KP tau-functions,  namely:
\begin{theorem}
\label{T-equivalence}
All polynomial  tau-functions of the KP hierarchy are, up to a   constant factor,  of the form
\begin{equation}
\label{tau-lambda1}
\tau_{\lambda;c}(t )=\det \left(s_{\lambda_j+i-j} (t_1+c_{1j},t_2+c_{2j},t_3+c_{3j}, \ldots)
\right)_{1\le i,j\le \ell},
\end{equation}
where $\lambda=(\lambda_1, \lambda_2,\ldots ,\lambda_\ell)\in Par_\ell$,  $\ell\ge 0$,   and  
$c= (c_{ij})$ is a $(\lambda_1+\ell-1)\times \ell$ matrix,   with  $c_{ij}\in\mathbb C$  and $c_{ij}=0$ for $i>\lambda_j-j+\ell$.
\end{theorem}

It is natural to call \eqref{tau-lambda1}   the generalized Jacobi-Trudi formula for polynomial KP tau-functions.
Similarly, the  polynomial KP tau-functions can be given in the {\it  Giambelli form} (see also \cite{EH}, Corollary 2.1, or \cite{HL}, \cite{HO}). Namely,  we can restate Theorem \ref{T-equivalence} in the following form.
\begin{theorem}
\label{T-Giambelli}
All polynomial  tau-functions of the KP hierarchy are, up to a  constant factor,  of the form
\begin{equation}
\label{tau-lambda2}
\tau_{\lambda;c,d}(t)=\det \left(\chi_{(a_i|b_j)} (t_1+c_{1i},t_2+c_{2i},  \ldots;t_1+d_{1j},t_2+d_{2j}, \ldots)
\right)_{1\le i,j\le k},
\end{equation}
where 
$
\chi_{(a |b)}(s;t) $ is given by \eqref{chi}.
Here $\lambda=(a_1,a_2,\ldots, a_k|b_1,b_2,\dots, b_k)$
 is  the Frobenius notation for $\lambda\in Par$,  and  
 $c=(c_{ij})$ a $(a_1+b_1+1)\times k$ matrix and 
 $d=(d_{ij})$ a $b_1\times k$ matrix,   with  $c_{ij}, d_{ij}\in\mathbb C$  and 
 $c_{ij}=0$ (resp  $d_{ij}=0$) for $i>a_j+b_1+1$ (resp. $i>b_j$).
\end{theorem}

Since the Giambelli formula  for Schur polynomials \eqref{Giam} is obtained from \eqref{tau-lambda2} by substituting 
$c_{ij}=d_{ij}=0$  for all $1\le i\le k$ and $j=1,2,\ldots$,
we call formula \eqref{tau-lambda2} {\it the generalized Giambelli formula for polynomial KP tau-functions}.

\begin{remark}
Let us compare our result \eqref{tau-lambda2} with the formulas in Section 2 of \cite{HO}.   We only consider the case  
$|\lambda; n\rangle$ for $n=0$, all the other cases can be obtained by a shift of indices.  
Formula \eqref{tau-lambda2} coincides with 
(2.49) of \cite{HO}, i.e. 
\[
s_{\lambda;0}({\bf t}(\tilde A))=\pi_{\lambda ;0}(g)({\bf t})=\langle 0|\exp \left(\sum_{i=1}^\infty t_i\alpha_i\right)\hat g f_\lambda,
\]
with
\begin{equation}
\label{flambda}
f_\lambda=e_{\lambda_1-\frac12}\wedge  e_{\lambda_2-\frac32}\wedge\cdots\wedge e_{\lambda_\ell-\ell+\frac{1}2}\wedge e_{-\ell-\frac12}\wedge e_{-\ell-\frac32}\wedge\cdots , \quad \lambda\in Par_\ell.
\end{equation}
The  action of $\hat g$ on the fermions  is  given by $\hat g\psi_{-\ell-\frac12}^+\hat g^{-1}=\sum_i g_{i\ell}(\tilde A)\psi_{-i-\frac12}^+$ and $\hat g\psi_{\ell+\frac12}^-\hat g^{-1}=
\sum_i g^{-1}_{\ell,i}(\tilde A)\psi_{i+\frac12}^-$.
For $\lambda=(a_1,a_2,\ldots, a_k|b_1,b_2,\dots, b_k)$ the   coefficients $g_{i\ell}(\tilde A)$ and $g^{-1}_{i\ell}(\tilde A)$ for $i,\ell\in\mathbb{Z}$ are as follows:
\[
\begin{aligned}
g_{i\ell}(\tilde A)&=\delta_{i\ell}&\ \mbox{for }0>\ell\ne  -b_1-1, \ldots,-b_k-1, \\
g_{i,a_j}(\tilde A)&=s_{a_j-i}(c_j)
&\ \mbox{for }j=1, \ldots, k,\\
g_{i,\ell}^{-1}(\tilde A)&=\delta_{i\ell}&\ \mbox{for }0\le i\ne a_1, \ldots, a_k,\\
g_{-b_j-1,\ell}^{-1}(\tilde A)&=s_{ \ell+b_j+1}(-d_j)&\ \mbox{for }j=1, \ldots, k,
\end{aligned}
\]
where  $c_i=(c_{1i},c_{2i},c_{3i},\ldots, c_{a_i+b_1+1,i})$,
 $d_i=(d_{1i},d_{2i},d_{3i},\ldots,d_{b_i,i})$ for $i=1,\ldots,k $,  with  $c_{ji}, d_{ji}\in\mathbb C$ arbitrary.
\end{remark}

Before giving the proof of Theorem \ref{T-Giambelli},  we will first state and prove a lemma,  and discuss the ideas of the proof of Theorem \ref{T-equivalence} (\cite{KvdLmodKP}, Theorem 16).
\begin{lemma}
\label{lemma1}
\[
\begin{aligned}
(a)&\qquad\exp (H(t)) | m\rangle= | m\rangle, \quad m\in\mathbb Z,\\
(b)&\qquad
(\exp H(t))\psi^{\pm }(z)\exp(-H(t))= \psi^{\pm }(z)\exp\left(\pm \sum_{k>0} t_k z^k\right)
, \\
(c)&\qquad
\langle (\exp H(t))\psi^+(y)\psi^{-}(z) \rangle=i_{y,z}\frac{1}{y-z} \exp\left( \sum_{k>0} t_k (y^k-z^k)\right),\\
(d)&\qquad
\langle (\exp H(t))\psi^+_{-i-\frac12}\psi^{-}_{-j-\frac12}\rangle= \sum_{\ell=0}^{j} s_{i+\ell+1}(t)s_{j-\ell}(-t),
 \\
(e)&\qquad
\langle (\exp H(t))\psi^+_{-i-\frac12}\psi^{-}_{-j-\frac12}\rangle=(-1)^j \chi_{(i|j)}(t;t)=(-1)^j s_{(i+1,1^j)}(t).
\end{aligned}
\]
\end{lemma}
{\bf Proof.} (a) follows from the fact that all $\alpha_k|0\rangle = 0$ for all $k>0$.\\
(b)   follows from  the fact that 
 $[ \alpha_{k}, \psi^{\pm }(z)]=\pm  z^k\psi^{\pm }(z)$.
 \\
 (c)  follows from (a),  (b) and the fact that $\langle \psi^+(y)\psi^{-}(z) \rangle=i_{y,z}\frac{1}{y-z} $.
 \\
 (d) follows by taking the coefficient of $y^{i}z^j$ in (c).  
 \\
 Finally,  (e) follows from (d),  the equality 
 \[
 \psi^+_{-i-\frac12}\psi^{-}_{-j-\frac12}|0\rangle=(-1)^je_{i+\frac12}\wedge e_{-\frac12}\wedge e_{-\frac32}\wedge\cdots\wedge
 e_{\frac12-j}\wedge e_{-j-\frac32}\wedge e_{-j-\frac52}\wedge\cdots, 
 \]
 and the fact  that (cf. \cite{KvdLmodKP},  Section 6)
 \[
 \sigma(e_{i+\frac12}\wedge e_{-\frac12}\wedge e_{-\frac32}\wedge\cdots\wedge
 e_{\frac12-j}\wedge e_{-j-\frac32}\wedge e_{-j-\frac52}\wedge\cdots)= s_{(i+1,1^j)}(t). 
 \]
\hfill$\square$
\\

Since a tau-function is independent of the charge $m$, see Remark \ref{remarktau},  we may assume that $m=0$.   Then by the Bruhat decomposition of $GL_\infty$, the $GL_\infty $  orbit ${\cal O}_0$ of $|0\rangle$ is   the disjoined union of Schubert cells:
\begin{equation}
{\cal O}_0=\bigcup_{\lambda\in Par }R(U) f_\lambda,
\end{equation}
where  $f_\lambda$ is given by \eqref{flambda}
and $U$ is the subgroup of $GL_\infty $ consisting of all  upper-triangular matrices, with $1$ on the diagonal. 

Note that the elements $f_\lambda$  form a basis of $F^{(0)}$,  and (see \cite{KPeterson},Theorem 4.1)
$\sigma(f_\lambda)=s_{\lambda}(t)$.

Now let $A= (a_{ij})\in U$,  then  
$
R(A) |{-\ell}\rangle= |{-\ell}\rangle$, for any $\ell\in\mathbb Z$,  since $|-\ell\rangle$ is the highest weight vector of $F^{(-\ell)}$.
Thus 
\[
\begin{aligned}
R(A)f_\lambda&=Ae_{\lambda_1-\frac12}\wedge  Ae_{\lambda_2-\frac32}\wedge\cdots\wedge Ae_{\lambda_\ell-\ell+\frac{1}2}\wedge e_{-\ell-\frac12}\wedge e_{-\ell-\frac32}\wedge\cdots\\
&=w_{\lambda_1-\frac12}\wedge  w_{\lambda_2-\frac32}\wedge\cdots\wedge w_{\lambda_\ell-\ell+\frac{1}2}
\wedge e_{-\ell-\frac12}\wedge e_{-\ell-\frac32}\wedge\cdots
 ,
 \end{aligned}
\]
where 
\[
w_{\lambda_j-j+\frac{1}2}=e_{\lambda_j-j+\frac{1}2}+\sum_{i\le  \lambda_j-j-\frac12} a_{ij}e_i.
\]
In fact we may obviously  assume that 
\begin{equation}
\label{w_j}
w_{\lambda_j-j+\frac{1}2}=e_{\lambda_j-j+\frac{1}2}+\sum_{i= \frac12-\ell}^{ \lambda_j-j-\frac12} a_{i,\lambda_j-j+\frac{1}2}e_j.
\end{equation}
Hence
\begin{equation}
\label{RR1}
R(A)f_\lambda=
 w_{\lambda_1-\frac12}^+ w_{\lambda_2-\frac32}^+\cdots w_{\lambda_\ell-\ell+\frac{1}2}^+|-\ell\rangle,
\end{equation}
where  
\begin{equation}
\label{w_j^+}
w_{\lambda_j-j+\frac{1}2}^+= \psi^+_{ \frac{2j-1}2-\lambda_j}+ \sum_{i\ge j+\frac12 -\lambda_j } a_{-i,\lambda_j-j+\frac{1}2}\psi_i^+,
\end{equation}
 and 
 \[
 w_{\lambda_j-j+\frac{1}2}^+={\rm Res }_z a_j(z)\psi^+(z),\quad \mbox{with } a_j(z)=z^{  j-\lambda_j-1}+\sum_{i\ge j+\frac12 -\lambda_j } a_{-i,\lambda_j-j+\frac{1}2}z^{i-\frac12}.
 \]
Now we can find constants $c_{\lambda_j-j+\frac{1}2}=(c_{1,\lambda_j-j+\frac{1}2}, c_{2,\lambda_j-j+\frac{1}2},\ldots)$   such that 
$a_{\lambda_j-j+\frac{1}2}(z)=z^{  j-\lambda_j-1}\exp \sum_{i=1}^\infty c_{i,\lambda_j-j+\frac{1}2}z^i$,  thus
\begin{equation}
\label{w_j^+2}
 w_{\lambda_j-j+\frac{1}2}^+={\rm Res}_z z^{  j-\lambda_j-1}\exp (\sum_{i=1}^\infty c_{i,\lambda_j-j+\frac{1}2}z^i )\psi^+(z).
\end{equation}
Finally,  using \eqref{exptau}, we find that (see  \cite{KRvdL},  Section 3)
\[
\begin{aligned}
\sigma&(w_{\lambda_1-\frac12}^+ w_{\lambda_2-\frac32}^+\cdots w_{\lambda_\ell-\ell+\frac{1}2}^+|-\ell\rangle)= 
\langle 0|\exp (H(t))w_{\lambda_1-\frac12}^+ w_{\lambda_2-\frac32}^+\cdots w_{\lambda_\ell-\ell+\frac{1}2}^+|-\ell\rangle\\
&={\rm Res}_{z_1}\cdots {\rm Res}_{z_\ell}z_1^{-\lambda_1}\cdots z_\ell^{\ell-\lambda_\ell- 1}
\exp \left(\sum_{j=1}^\ell\sum_{i=1}^\infty c_{i,\lambda_j-j+\frac{1}2}z_j^i \right)
\langle 0|\exp (H(t))
\psi^+(z_1)\cdots\psi^+(z_\ell)|-\ell\rangle\\
&={\rm Res}_{z_1}\cdots {\rm Res}_{z_\ell}
z_1^{-\lambda_1-\ell}z_2^{1-\lambda_2-\ell}\cdots z_\ell^{-\lambda_\ell- 1}
\prod_{1\le j<k\le \ell} (z_j-z_k)
\exp \left(\sum_{j=1}^\ell\sum_{i=1}^\infty (t_i+c_{i,\lambda_j-j+\frac{1}2})z_j^i \right),
\end{aligned}
\]
which is equal to \eqref{tau-lambda1}, where one has to replace $c_j$ by $c_{\lambda_j-j+\frac{1}2}$.  
In the above calculations we assume that $|z_1|>|z_2|>\cdots>|z_\ell|$.
It is clear from the above generalized Jacobi-Trudi formula \eqref{tau-lambda1}, that the concstants $c_{ij}$ for $i>\lambda_j-j+\ell$ do not appear there,  so  we can choose them all equal to 0,   which gives the restriction that $c_j \in\mathbb{C}^{\lambda_j-j+\ell}$
\\
\  

{\bf Proof of Theorem \ref{T-Giambelli}.}  
We first  rewrite  \eqref{flambda}.  Since 
\[
\lambda=(\lambda_1,
\lambda_2,\cdots ,\lambda_\ell)= (a_1,a_2,\ldots, a_k|b_1,b_2,\dots, b_k),
\]
we find that
\[
\begin{aligned}
f_\lambda=& 
\psi^+_{\frac12-\lambda_1 }\psi^+_{\frac32- \lambda_2 }\cdots \psi^+_{\ell-\frac{1}2-\lambda_\ell}
\psi^-_{\frac12-\ell}\psi^-_{\frac32-\ell}\cdots \psi^-_{-\frac12}|0\rangle\\
=&
\psi^+_{-a_1-\frac12 }\psi^+_{-a_2 -\frac12}\cdots \psi^+_{-a_k -\frac12}
 \psi^+_{k+\frac{ 1}2-\lambda_{k+1}}
\cdots \psi^+_{\ell-\frac{1}2-\lambda_\ell}
\psi^-_{\frac12-\ell}\psi^-_{\frac32-\ell}\cdots \psi^-_{-\frac12}|0\rangle.
\end{aligned}
\]
Observe   that $a_j=\lambda_j-j\ge 0$,  for $j=1,\ldots ,k$ and $\lambda_j-j<0$ for $j>k$.
We now move  all $\psi^+_{j-\frac{ 1}2-\lambda_j}$ with $j>k$ to the right of all $\psi^-_{-i}$ with $0<i<\ell$.
This has the effect that one  removes all $\psi^-_{-i}$, for which $i$ is equal to one of the ${j-\frac{ 1}2-\lambda_j}$ for $j>k$. This gives that $f_\lambda $
is equal,  up to a sign, which we may ignore, to
\begin{equation}
f_\lambda=\psi^+_{-a_1-\frac12 }\psi^+_{-a_2 -\frac12}\cdots \psi^+_{-a_k -\frac12}
\psi^-_{-b_1-\frac12 }\psi^-_{-b_2 -\frac12}\cdots \psi^-_{-b_k -\frac12}|0\rangle.
\end{equation}
Next we   calculate $R(A)f_\lambda$ for $A\in U$.
Clearly $A \psi^+_{-a_j -\frac12}=A\psi^+_{j-\frac{ 1}2-\lambda_j}=w_{{\lambda_j-j+\frac{1}2}}^+$ for  $j=1,\ldots, k$,  where $w_{{\lambda_j-j+\frac{1}2}}^+=w_{a_j+\frac12}^+$ is as in \eqref{w_j^+},  and thus  is equal \eqref{w_j^+2}.

	Let $w_{{-b_j-\frac12}}^-= A\psi^-_{-b_j-\frac12}$,  then there exist  constants
	$c_{-b_j-\frac12}=(c_{1,-b_j-\frac12},c_{2,-b_j-\frac12},\ldots)$, such that 
\[
\begin{aligned}
w_{-b_j-\frac12}^-=& \psi^-_{-b_j-\frac12}+\sum_{i\ge  \frac12-b_j} c^-_{i,-b_j-\frac12} \psi^-_i \\
=&{\rm Res}_z ( z^{-b_j-1}+\sum_{i\ge  \frac12-b_j} c^-_{i,-b_j-\frac12} z^{i-\frac12}
)\psi^-(z).
\end{aligned}
\]
In a similar way as for the $w^+_{a_i+\frac12}$,  there exist constants $d_{-b_j-\frac12}=(d_{1,-b_j-\frac12},d_{2,-b_j-\frac12},\ldots)$, such that
\begin{equation}
\label{w_b^-}
w_{-b_j-\frac12}^-={\rm Res}_z  z^{-b_j-1}
\exp (-\sum_{i=1}^\infty d_{i,-b_j-\frac12}z^i) \psi^-(z).
\end{equation}
Using \eqref{w_j^+2} and \eqref{w_b^-}, 
the relation between the constants $c_i$ and $d_{-j}$ is as follows.  Since $(w^+_i,w^-_{-j})=\delta_{i,-j}$, we find that
\[
\begin{aligned}
\delta_{i,-j}=&
{\rm Res}_y {\rm Res}_z 
y^{ -i-\frac12}
z^{-j-\frac12}
\exp (\sum_{a=1}^\infty c_{a,j}
 y^a)
\exp (-\sum_{a=1}^\infty d_{a,-j}z^a) 
(\psi^+(y),\psi^-(z))\\
=&{\rm Res}_y {\rm Res}_z 
y^{ -i-\frac12}
z^{-j-\frac12}
\exp (\sum_{a=1}^\infty c_{a,j}y^a-d_{a,-k}z^a
  )
 \delta(y-z)\\
=&{\rm Res}_y {\rm Res}_z 
z^{ -i-j-1}
\exp (\sum_{a=1}^\infty (c_{a,j}-d_{a,-k})z^a
  )
 \delta(y-z)\\
 =&  {\rm Res}_z 
z^{ -i-j-1}
\exp (\sum_{a=1}^\infty (c_{a,j}-d_{a,-k})z^i
  )
  \\
  =& s_{i+j}(c_i-d_{-j}).
\end{aligned}
\]
This means that we have a possible restiction on the constants  $d_{-b_j-\frac12}$ for $j=1,\ldots,k$, viz.
\[
s_{a_i+b_j+1}(c_{a_i+\frac12}-d_{-b_j-\frac12})=0\quad \mbox{for }   1\le i,j\le k.
\]
Stated differently,
\begin{equation}
\label{restrict}
d_{a_i+b_j+1,-b_j-\frac12}=s_{a_i+b_j+1}
(c_{1,a_i+\frac12}-d_{1,-b_j-\frac12}, \ldots ,c_{a_i+b_j,a_i+\frac12}-d_{a_i+b_j,-b_j-\frac12},c_{a_i+b_j+1,a_i+\frac12}).
\end{equation}

Finally we calculate 
\[
\begin{aligned}
\sigma(R(A)f_\lambda)=&\langle 0|(\exp H(t))w_{a_1+\frac12}^+\cdots 
w_{a_k+\frac12}^+w_{-b_1-\frac12}^-\cdots 
w_{-b_k-\frac12}^-|0\rangle .
\end{aligned}
\]
Using Wick's theorem, this is equal,  up to a sign, to 
$$\det(\langle (\exp H(t)))w_{a_i+\frac12}^+w_{-b_j-\frac12}^-\rangle)_{1\le i,j\le k}.$$
Now,  using \eqref{w_j^+2},  \eqref{w_b^-},  and  Lemma \ref{lemma1},  we find that  in the domain $|y|>|z|$ we have 
\[
\begin{aligned}
&\langle (\exp H(t))w_{a_i+\frac12}^+w_{-b_j-\frac12}^-\rangle=
\\
&{\rm Res}_y{\rm Res}_z y^{ - a_i-1}z^{-b_j-1}\exp (\sum_{r=1}^\infty c_{r,a_i+\frac{ 1}2}y^r
-d_{r,-b_j-\frac12}z^r)
\langle(\exp H(t))\psi^+(y)
  \psi^-(z)\rangle=\\
 & {\rm Res}_y{\rm Res}_z y^{  -a_i-1}z^{-b_j-1}
  i_{y,z}\frac{1}{y-z} \exp( \sum_{r>0} (t_r + c_{r,a_i+\frac{ 1}2})
  y^r-((t_r +d_{r,-b_j-\frac12})z^r)=\\
  & {\rm Res}_y{\rm Res}_z  \sum_{p,q,r=0}^\infty y^{ p- r-a_i-2}z^{q+r-b_j-1}
    s_p(t+c_{a_i+\frac{ 1}2})s_q(-(t+d_{-b_j-\frac12}))=
  \\
   & \sum_{r=0}^\infty s_{r+a_i+1}(t+c_{a_i+\frac{ 1}2})s_{b_j-r}(-(t+d_{-b_j-\frac12}))=
  \\
  &(-1)^{b_j}\chi_{(a_i|b_j)}(t+  c_{a_i+\frac{ 1}2},;t +d_{-b_j-\frac12}).
  \end{aligned}
\]
Hence the tau-function $\sigma(R(A)f_\lambda)$ is equal to \eqref{tau-lambda2}, where we replace the constants $c_i$ (resp.  $d_i$) by $c_{a_i+\frac{ 1}2}$ (resp. $d_{-b_i -\frac12}$).

Note that the constants $d_{m,-b_j-\frac12}$ that appear in $\chi_{(a_i|b_j)}(t+  c_{a_i+\frac{ 1}2},;t +d_{-b_j-\frac12})$ are $d_{1,-b_j-\frac12},\ldots, d_{b_j,-b_j-\frac12}$, and the $d_{m,-b_j-\frac12}$, with $m>b_j$ do not appear.
However, looking  at the restriction \eqref{restrict}, we see that the first dependence of the constants 
$d_{m,-b_j-\frac12}$ on $d_{r,-b_j-\frac12}$  with $r<m$, and on the  $c_{n, a_i+\frac12}$  with $i=1,\ldots k$, $n=1,2,\ldots$,  is for $m=a_k+b_j+1>b_j$.  But since the only coefficients that appear in the tau-function are the $d_{m,-b_j-\frac12}$   with $m\le b_j$,  the restriction \eqref{restrict} is void. 

If we look at which elementary Schur polynomials appear in $\chi_{(a_i|b_j)}(t;t')$ we see that the constants $c_{nj}$(resp. $d_{nj}$) with $n>a_j+b_1+1=\lambda_j-j+\ell$ (resp.   $n>b_j$)  do not appear.
\hfill$\square$

\section{More general tau-functions and the wave function of the KP hierarchy}
\label{S4}
Following   \cite{KRvdL}, 
we can construct  generating functions of tau-functions using the Jacobi-Trudi formalism.
Namely,  we consider,  instead of $R(A)f_\lambda$, the element
\[
\psi^+(z_1)\psi^+(z_2)\ldots\psi^+(z_\ell)\psi^-_{\frac12-\ell}\psi^-_{\frac32-\ell}\cdots \psi^-_{-\frac12}|0\rangle,
\]
where we assume $|z_i|>|z_{i+1}|$ for all $i=1,\ldots,\ell-1$.
Let 
\begin{equation}
\label{Tdef}
\begin{aligned}
T(z_1,\ldots,z_\ell)&=
\prod_{1\le j<k\le \ell} (z_j-z_k)
\exp \left(\sum_{j=1}^\ell\sum_{n=1}^\infty t_nz_j^n\right)\\
&=\det\left(z_i^{j-1}\exp \left( \sum_{n=1}^\infty t_nz_i^n\right)\right)_{1\le i,j\le \ell}
.
\end{aligned}
\end{equation}
Then the corresponding tau-function  is equal to
\begin{equation}
\label{gentau}
\begin{aligned}
\langle 0|(\exp H(t))
\psi^+(z_1)\cdots\psi^+(z_\ell)|-\ell\rangle 
&= z_1^{-\ell}\cdots z_\ell^{-\ell}T(z_1, \ldots z_\ell)
.
\end{aligned}
\end{equation}
If we now take consequtive  residues of $ T(z_1,\ldots,z_\ell)\prod_{i=1}^\ell a_i(z_i)$,  where 
$a_i( z)$ are  some    Laurent series  in $z$,   we obtain a tau-function:
\begin{equation}
\label{taugentau}
\tau(t)={\rm Res}_{z_1}\cdots {\rm Res}_{z_\ell }\, T(z_1,\ldots,z_\ell)\prod_{i=1}^\ell a_i(z_i).
\end{equation}
These expressions are not polynomial    in general.
To obtain the polynomial tau-function  of the previous section,  we take 
\begin{equation}
\label{a_j}a_j(z)= z^{  j-\ell-\lambda_j-1}\exp (\sum_{i=1}^\infty c_{i,\lambda_j-j+\frac{1}2}z^i ).
\end{equation}
But $T(y_1,\ldots,y_\ell)$   is also a tau-function, for this one chooses $a_i(z_i)=\delta(z_i-y_i)$.
One can even construct the wave function \eqref{wave1} in this way.
Assume $\tau(t)$ is given by \eqref{taugentau},  then  
\begin{equation}
\label{30x}
\tau(t)w^+(t,z)=\langle 1|(\exp H(t))\psi^+(z)w_1^+\cdots w_\ell^+|-\ell\rangle ,
\end{equation}
where,  by   \eqref{w_j^+}   , $w^+_j= {\rm Res}_{z_j} a_j(z_j)\psi^+(z_j)$,  so that 
\[
\langle 1|(\exp H(t))\psi^+(z)\psi^+(z_1)\cdots\psi^+(z_\ell)|-\ell\rangle= z^{-\ell}z_1^{-\ell}\cdots z_\ell^{-\ell}T(z,z_1,  \ldots, z_\ell).
\]
Thus 
\begin{equation}
\label{wavegentau}
\tau(t)w^+(t,z)= z^{-\ell} {\rm Res}_{z_1}\cdots {\rm Res}_{z_\ell }\, T(z,z_1,
\ldots,z_\ell)\prod_{i=1}^\ell a_i(z_i),
\end{equation}
and, using \eqref{taugentau},  the wave function is equal to 
\begin{equation}
\label{wave1a}
w^+(t,z)=z^{-\ell}
\frac{
 {\rm Res}_{z_1}\cdots {\rm Res}_{z_\ell }\, T(z,z_1,
\ldots,z_\ell,)
\prod_{i=1}^\ell a_i(z_i)
}
{{\rm Res}_{z_1}\cdots {\rm Res}_{z_\ell }\, T(z_1,\ldots,z_\ell)\prod_{i=1}^\ell a_i(z_i)}.
\end{equation}
Taking all $a_j(z)$ as in \eqref{a_j},   with $ c_{\lambda_j-j+\frac{1}2}$,  replaced by $c_j$,  we obtain, as denominator  of $w^+(t,z)$, the polynomial  $\tau_\lambda (t)$ given by \eqref{tau-lambda1}.     
Using \eqref{gentau} and \eqref{30x}, by Wick's theorem we obtain the numerator of \eqref{wave1}.
Thus
\begin{equation}
w^+(t,z)=\frac1{\tau_\lambda(t)}
\det
\begin{pmatrix}
e^{\sum_{n} t_nz^n}&z^{-1}e^{\sum_{n} t_nz^n}&z^{-2}e^{\sum_{n} t_nz^n}&\cdots &z^{-\ell}e^{\sum_{n} t_nz^n}\\
s_{\lambda_1-1}(t+c_1)&s_{\lambda_1}(t+c_1)&s_{\lambda_1+1}(t+c_1)& \cdots&s_{\lambda_1+\ell}(t+c_1)\\
s_{\lambda_2-2}(t+c_2)&s_{\lambda_2-1}(t+c_2)&s_{\lambda_2}(t+c_2)&\cdots&s_{\lambda_2+\ell-1}(t+c_1)\\
\vdots&\vdots&\vdots&\ddots&\vdots\\
s_{\lambda_\ell-\ell}(t+c_\ell)&s_{\lambda_\ell-\ell+1}(t+c_\ell)&s_{\lambda_\ell}(t+c_\ell)&\cdots&s_{\lambda_\ell}(t+c_\ell)
\end{pmatrix}
.
\end{equation}
We call this the the Jacobi-Trudi formula for  the  wave function related to a polynomial tau-functions.
Note that this implies that $\tau(t)w^+(t,z)$ is also a tau-function.

We now want to do a similar thing using the Giambelli formalism.   For this we consider  for $|y_i|>|y_{i+1}$, $|z_i|>|z_{i+1}|$ for all $i=1,\ldots, k-1$, and $|y_i|>|z_j|$ for $1\le i,j\le k$.    Consider the element
\[
\psi^+(y_1)\psi^+(y_2)\ldots\psi^+(y_k)
\psi^-(z_1)\psi^-(z_2)\ldots\psi^-(z_k)
 |0\rangle.
\]
We want to calculate the corresponding tau-function, i.e. 
\[
\langle 0|\exp((H(t))
\psi^+(y_1)\psi^+(y_2)\ldots\psi^+(y_k)
\psi^-(z_1)\psi^-(z_2)\ldots\psi^-(z_k)
 |0\rangle.
\]
Using Wick's theorem, since we have fermions,  we obtain a Pfaffian,
\[
Pf \begin{pmatrix}
0&(\langle 0|\exp((H(t))
(\psi^+(y_i) 
\psi^-(z_j) 
 |0\rangle)_{ij}\\
 -(\langle 0|\exp((H(t))
(\psi^+(y_j) 
\psi^-(z_i) 
 |0\rangle)_{ij}&0
\end{pmatrix}_{1\le i,j\le k},
\]
wich is   equal, up to   sign, to 
\begin{equation}
\label{gentauS}
S(y_1,\ldots,y_k;z_1,\ldots, z_k)=\det \left(  \frac{\exp \left( \sum_{n=1}^\infty t_n(y_i^n-z_j^n)\right)}{y_i-z_j}\right)_{1\le i,j\le k}.
\end{equation}
To obtain the polynomial tau-functions of the previous section,  we take
\begin{equation}
\label{a,b}
a_i(y)=y^{ - a_i-1}\exp (\sum_{r=1}^\infty c_{r,a_i+\frac{ 1}2}y^r
 ),\quad 
b_i(z)=z^{-b_i-1}\exp (-\sum_{r=1}^\infty  d_{r,-b_i-\frac12}z^r),
\end{equation}
and let 
\begin{equation}
\label{taugentau2}
\tau(t)={\rm Res}_{y_1}\cdots {\rm Res}_{y_k }{\rm Res}_{z_1}\cdots {\rm Res}_{z_k }S(y_1,\ldots,y_k;z_1,\ldots, z_k)  \prod_{i=1}^k a_i(y_i)b_j(z_j).
\end{equation}
The question we want to solve is:  can we also express the wave function in this way using  formula \eqref{gentauS}?
Recall that the wave function,  multiplied by $ \tau(t)$ can be calculated, by multiplying  $\sigma^{-1}(\tau(t))$ by $\psi^+(z)$ and then calculating $\sigma$ of this.  In other words
\[
\tau(t)w^+ (t,z)=\langle - 1|(\exp H(t))\psi^+(z)R(G)|0\rangle.
\]
So we  to calculate, for $|z|>|y_1|>\ldots |y_k|>|z_{1}1>\ldots>|z_{k}|$:
\[
\begin{aligned}
\langle & 1|(\exp H(t))\psi^+(z)\psi^+(y_1)\psi^+(y_2)\ldots\psi^+(y_k)
\psi^-(z_{1})\psi^-(z_{2})\ldots\psi^-(z_{k})
 |0\rangle\\
 &=\langle 0|\psi_{\frac12}^{-} (\exp H(t))\psi^+(z)\psi^+(y_1)\psi^+(z_2)\ldots\psi^+(y_k)
\psi^-(z_{1})\psi^-(z_{2})\ldots\psi^-(z_{k})
 |0\rangle\\
 \\
& = \exp \left(  \sum_{n=1}^\infty t_n z^n\right)\exp \left(\sum_{i=1}^k \sum_{n=1}^\infty t_n(y_i^n-z_{i}^n)\right)\times\\
 &\qquad\qquad \langle 0|\psi_{\frac12}^{-}
 \psi^+(z)\psi^+(y_1)\psi^+(y_2)\ldots\psi^+(y_k)
\psi^-(z_{1})\psi^-(z_{2})\ldots\psi^-(z_{k})
 |0\rangle\\
 &=\pm \det
 \begin{pmatrix}
 1&\frac1{z-z_{1}}&\frac1{z-z_{2}}&\cdots &\frac1{z-z_{-k}}\\
 1&\frac1{y_{ 1}-z_{ 1}}& \frac1{y_{  1-}z_{2}}&\ldots &\frac1{y_1-z_{k}}\\
 1&\frac1{y_{ 2}-z_{1}}& \frac1{y_{2}-z_{2}}&\ldots &\frac1{y_{2}-z_{k}}\\
 \vdots&\vdots&\vdots&\ddots&\vdots\\
 1&\frac1{y_{k}-z_{ 1}}& \frac1{y_{k}-z_{ 2}}&\ldots &\frac1{y_{k}-z_{k}}
 \end{pmatrix}
 \exp \left(\sum_{i=1}^k \sum_{n=1}^\infty t_n(y_i^n-z_{i}^n)\right)\exp \left(  \sum_{n=1}^\infty t_n z^n\right)
 \\
 &=\pm\det \begin{pmatrix}
 e^{\sum_n  t_nz^n}&\frac{e^{\sum_n  t_n(z^n-z_{1}^n)}}{z-z_{1}}&\frac{e^{\sum_n  t_n(z^n-z_{2}^n)}}{z-z_{2}}&\cdots &\frac{e^{\sum_n  t_n(z^n-z_{k}^n)}}{z-z_{k}}\\
 e^{\sum_n  t_ny_{1}^n}&\frac{e^{\sum_n  t_n(y_{1}^n-z_{1}^n)}}{y_{1}-z_{1}}&\frac{e^{\sum_n  t_n(y_1^n-z_{2}^n)}}{y_1-z_{2}}&\cdots &\frac{e^{\sum_n  t_n(y_1^n-z_{k}^n)}}{y_1-z_{k}}\\
 e^{\sum_n  t_ny_{2}^n}&\frac{e^{\sum_n  t_n(y_{2}^n-z_{1}^n)}}{y_{2}-z_{1}}&\frac{e^{\sum_n  t_n(y_2^n-z_{2}^n)}}{y_2-z_{2}}&\cdots &\frac{e^{\sum_n  t_n(y_2^n-z_{k}^n)}}{y_2-z_{k}}\\
\vdots&\vdots&\vdots&\ddots&\vdots\\
 e^{\sum_n  t_ny_{k}^n}&\frac{e^{\sum_n  t_n(y_{k}^n-z_{1}^n)}}{y_{k}-z_{1}}&\frac{e^{\sum_n  t_n(y_k^n-z_{2}^n)}}{y_k-z_{2}}&\cdots &\frac{e^{\sum_n  t_n(y_k^n-z_{k}^n)}}{y_k-z_{k}}
 \end{pmatrix}\\
 &=\mp z_0S(z,y_1,\ldots, y_k;z_0,z_{1},\ldots, z_{k})|_{z_0=\infty}\qquad \mbox{for   } |z_0|>|z|
 .
\end{aligned}
\]
In the 3th equality, we used Wick's theorem and the observation that the Pfaffian, we obtain in this way, is of the form 
$$
Pf\begin{pmatrix}
0&A\\
A^T&0
\end{pmatrix}, 
$$ and  is   equal, up to a sign, to   $\det A$.
Hence,  the wave function is equal to 
\begin{equation}
\label{wave2}
w^+(t,z)=-
\frac{
{\rm Res}_{y_1}\cdots {\rm Res}_{y_k }{\rm Res}_{z_1}\cdots {\rm Res}_{z_k } z_0S(z,y_1,\ldots,y_k;z_0, z_1,\ldots, z_k) |_{z_0=\infty} \prod_{i=1}^k a_i(y_i)b_j(z_j)
}{
{\rm Res}_{y_1}\cdots {\rm Res}_{y_k }{\rm Res}_{z_1}\cdots {\rm Res}_{z_k }S(y_1,\ldots,y_k;z_1,\ldots, z_k)  \prod_{i=1}^k a_i(y_i)b_j(z_j)
}.
\end{equation}
Now, taking $a_i(z)$ and $b_i(z)$ as in \eqref{a,b},  thus replacing $c_{a_i+\frac12}$ by $c_i$ and  $d_{-b_i-\frac12}$ by $d_i$,   we obtain the wave function corresponding to the tau-function given by 
\eqref{tau-lambda2}:
\begin{equation}
\label{TTT}
\begin{aligned}
&w^+(t,z)=\frac1{\tau(t)}\exp\left(\sum_{n=1}^\infty t_nz^n\right)\times \\
&\det
\begin{pmatrix}
1&\chi_{(0|b_1)}([z^{-1}];t+d_1)&\chi_{(0|b_2)}([z^{-1}];t+d_2)&\cdots&\chi_{(0|b_k)}([z^{-1}];t+d_k)
\\
s_{a_1 }(t+c_1   )&\chi_{(a_1|b_1)}(t+c_1;t+d_1)&\chi_{(a_1|b_2)}(t+c_1;t+d_2)&\cdots&\chi_{(a_1|b_k)}(t+c_k;t+d_k)
\\

s_{a_2 }(t+c_2  )&\chi_{(a_2|b_1)}(t+c_2 ;t+d_1)&\chi_{(a_2|b_2)}(t+c_2;t+d_2)&\cdots&\chi_{(a_2|b_k)}(t+c_2;t+d_k)
\\
\vdots&\vdots&\vdots&\ddots&\vdots
\\
s_{a_k }(t+c_k)&\chi_{(a_k|b_1)}(t+c_k;t+d_1)&\chi_{(a_k|b_2)}(t+c_k;t+d_2)&\cdots&\chi_{(a_k|b_k)}(t+c_k;t+d_k)
\end{pmatrix}
.
\end{aligned}
\end{equation}
Here $[z^{-1}]=(\frac{z^{-1}}1,\frac{z^{-2}}2,\frac{z^{-3}}3,\ldots)$, $\chi_{(a|b)}$ is defined by \eqref{chi},  and
$\chi_{(0|b)}([z^{-1}];t)=(-1)^b\sum_{j=0}^b z^j s_{b-j}(-t)$.

\section{The  formulation of the CKP hierarchy}
\label{S5}
The  group $SP_\infty$,   the  corresponding Lie algebra $sp_\infty$  and its central extension  $c_\infty$ can be defined using the following bilinear form on  $\mathbb{C}^\infty$, see e.g. \cite{Kacbook},  Section 7.11:
\begin{equation}
\label{bilinform}
(e_i,e_j)_C=(-1)^{i+\frac12}\delta_{i,-j}.
\end{equation}
Then 
\[
\begin{aligned}
SP_\infty&=\{ G\in GL_\infty | (G(v),G(w))_C=(v,w)_C \ \mbox{for all } v,w\in \mathbb{C}^\infty\},\\
sp_\infty&=\{ g\in gl_\infty | (g(v),w)_C+(v,g(w))_C=0 \ \mbox{for all } v,w\in \mathbb{C}^\infty\},\\
c_\infty&=\{ g +\lambda K\in a_\infty | (g(v),w)_C+(v,g(w))_C=0 \ \mbox{for all } v,w\in \mathbb{C}^\infty\},\\
\end{aligned}
\]
The elements $C_{jk}=E_{-j,k}-(-1)^{j+k}E_{-k,j} $, with $j\ge k$
 form a basis of $sp_\infty$.

Note that 
\[
\begin{aligned}
r(C_{jk})& =\psi^+_{j }\psi^-_{k }-(-1)^{j+k} \psi^+_{k}\psi^-_{j }\quad\mbox{and } \hat r(C_{jk})& =:\psi^+_{j }\psi^-_{k }:+(-1)^{j+k} :\psi^-_{j }\psi^+_{k}:
.
\end{aligned}
\]
This suggests to define an automorphism  of the Clifford algebra $C\ell$:
\begin{equation}
\label{aut}
\iota_C(\psi_j^\pm)=(-1)^{j \pm \frac12}\psi^\mp_j.
\end{equation}
This induces via $\hat r$ and the observation that 
$
\iota_C(:\psi^+_{j }\psi^-_{k }:)=(-1)^{j+k}:\psi^-_{j }\psi^+_{k}:=-(-1)^{j+k}:\psi^+_{k }\psi^-_{j }:$
 the following involution on $a_\infty$:
$$\iota_C(E_{-j,k})=-(-1)^{j+k}E_{-k,j},\quad \iota_C(K)=K,$$
so that
$$c_\infty=\{g+\lambda K\in a_\infty |\iota_C(g)=g\}.$$

Let us study this automorphism and its consequences a bit better.
First of all 
$\iota_C({\rm Ann}_\pm |0\rangle)= {\rm Ann}_\mp  |0\rangle$ and thus 
$\iota_C({\rm Ann}\, |0\rangle)= {\rm Ann}\, |0\rangle$, 
which makes it possible to define the automorphism also on the module $F$,  namely define $\iota_C(|0\rangle)=|0\rangle$,  and the rest is induced by \eqref{aut}.
Since $\iota_C(\psi^\pm(z))=\pm \psi^\mp(-z)$,  we see that $\iota_C(F^{(m)})=F^{(-m)}$ and we deduce that
\[
\iota_C(\alpha(z))=\iota_C(:\psi^+(z)\psi^-(z)=:\psi^+(-z)\psi^-(-z):=\alpha(-z),
\]
and hence that $\iota_C(\alpha_k)=-(-1)^k\alpha_k$.
Moreover,
$\iota_C(|m\rangle)=(-1)^{\frac{m(m-1)}2}|-m\rangle$.
Using the bosonization $\sigma:F\to B$,  the automorphism $\iota_C$ defines an automorphism, wich we also denote by $\iota_C$,   of $B$. 
Clearly $\iota_C(1)=1$,  and we get  that the operators on $B$ satisfy
\[
\iota_C(t_i)=-(-1)^i t_i,\ \iota_C(\frac{\partial}{\partial t_i})= -(-1)^i \frac{\partial}{\partial t_i},\
\iota_C(q\frac{\partial}{\partial q})=-q\frac{\partial}{\partial q},\
\iota_C(q)=q^{-1}(-1)^{q\frac{\partial}{\partial q}}.
\]
In  particular, $F^{(0)}$ is $\iota_C$ invariant,  and  
\[
F_C^{(0)}=\{
f\in F^{(0)}|\, \iota_C(f)=f\}
\]
is an invariant subspace  for the representation $\hat r$,   restricted to
$c_\infty$.
Hence we want to look at $f\in{\cal O}_0$    that satisfies  the condition $\iota_C(f)=f$.
Let $v^\pm\in {\rm Ann}_\pm f$ for such an $f$,  then $v^\pm f=0$ and thus 
$\iota_C(v^\pm f )=\iota_C(v^\pm)f=0$.  Hence $\iota_C({\rm Ann}_\pm f)= {\rm Ann}_\mp  f$.
This makes it possible to define a skewsymmetric form on $\Psi$, given by
$\omega(v ,w)=(v  ,\iota_C (w))$;  more explicitly (cf.   \eqref{bilinform})
\begin{equation}
\label{bilomega}
\omega(\psi^\pm_i, \psi_j^\pm)=(\psi_i^\pm, (-1)^{j\pm\frac12}\psi_j^\mp)=(-1)^{i\mp\frac12}\delta_{i,-j},\qquad
\omega(\psi^+_i, \psi_j^-)=0.
\end{equation}
So we find that  ${\rm Ann}\, f$ is a maximal isotropic subspace of $\Psi=\Psi^+\oplus \Psi^-$, 
with respect to both the symmetric bilinear form $(\cdot ,\cdot )$ and the skewsymmetric $\omega(\cdot,  \cdot)$.
Since $f= G|0\rangle\in F^{(0)}_C$,  we find that ${\rm Ann}_\pm  |0\rangle\subset \Psi^\pm $ is maximal isotropic,  and therefore also  
${\rm Ann}_\pm f\subset \Psi^\pm$ is maximal isotropic.  Thus  ${\rm Ann}\, f$ (resp.  
${\rm Ann}_\pm f$)   is a Lagrangian subspace of $\Psi$ (resp.  $\Psi_\pm$).

Moreover, when applying the bosonization $\sigma$ to such an element $f$, we obtain a tau-function
$\tau(t)$,  wich satisfies
\begin{equation}
\label{tauC}
\tau(t_1,t_2,t_3,t_4,\ldots)=\iota_C(\tau(t_1,t_2,t_3,t_4,\ldots))=\tau(t_1,-t_2,t_3,-t_4,\ldots).
\end{equation}
\begin{remark}
Using formula (e) of Lemma \ref{lemma1} and \eqref{aut}, we find that 
$$
\begin{aligned}\iota_C(\chi_{(i|j)}(t,t))=&
(-1)^j\langle 0|(\exp H(t))\iota_C(\psi^+_{-i-\frac12}\psi^{-}_{-j-\frac12}|0\rangle)\\
=&(-1)^{i+1}\langle (\exp H(t))\psi^-_{-i-\frac12}\psi^{+}_{-j-\frac12}\rangle\\
=&(-1)^{i}\langle (\exp H(t))\psi^{+}_{-j-\frac12}\psi^-_{-i-\frac12}\rangle\\
=&\chi_{(j|i)}(t,t).
\end{aligned}
$$
Hence, using the Giambelli formula \eqref{Giam}, we find that $\iota_C(s_\lambda(t))=s_{\lambda'}(t)$, where $\lambda'$ is the conjugate partition of $\lambda$. Thus the only Schur polynomials that are invariant under $\iota_C$ are the ones for which $\lambda$ is self-conjugate. 
\end{remark}

Next apply  $1\otimes\iota_C$ or t$\iota_C\otimes 1$ to \eqref{mKP} for $m=0 $,  and assume that $\iota_C(f_0)=(f_0)$.
Then equation \eqref{mKP} turns into 
\begin{equation}
\label{CKP}
{Res}_{z}\,  \psi^\pm(z)f_0\otimes \psi^\pm(-z)f_0=0, \quad f_0\in F^{(0)}_C.
\end{equation}
\begin{remark}
Assume that $f_0$ satisfies the fermionic KP equation \eqref{mKP}, and that \eqref{CKP} holds  in  the - case. Then this gives that 
${\rm Ann}_+ f_0$ is maximal isotropic with respect to $\omega(\cdot,\cdot)$.   Namely,  let 
\[
f_0=v_1\wedge v_2\wedge v_3\wedge\cdots , 
\]
where we may assume, without loss of generality,   that $v_{j}=e_{-j+\frac12}$ for $j>>0$.
Let $(e_i,e_j)=\delta_{ij}$, then equation \eqref{CKP} gives that
\[
\begin{aligned}
0=&\sum_{i\in\frac12+\mathbb Z}(-1)^{i-\frac12}\psi^-_i f_0\otimes \psi^-_{-i} f_0
\\
=&
\sum_{k,\ell>0}\sum_{i\in\frac12+\mathbb Z}(-1)^{i+k+\ell-\frac12}(e_{-i},v_k)(e_{i},v_\ell)
v_1\wedge\cdots  v_{k-1}\wedge v_{k+1}\wedge\cdots \otimes
v_1\wedge\cdots  v_{\ell-1}\wedge v_{\ell+1}\wedge\cdots  .
\end{aligned}
\]
Since the vectors $v_1,v_2,\ldots$ are linearly independent, we find that 
$$\sum_{i\in\frac12+\mathbb Z}(-1)^{i+\frac12}(e_i,v_k)(e_{-i},v_\ell)=0\ \mbox{ for all }k,\ell=1,2,\ldots,$$
which means that 
(cf. \eqref{1.19}) $(v_k,v_\ell)_C=0$ for all $k,\ell=1,2,\ldots$.  This, and the fact that     $v_{j}=e_{-j+\frac12}$ for $j>>0$, 
gives   that ${\rm Ann}_+ f_0$ is maximal isotropic with respect to $\omega(\cdot,\cdot)$.
\end{remark}

Letting $\tau(t)=\sigma(f_0)$,  gives, by \eqref{bosonic-mKP}, the following CKP hierarchy of equations on the tau-function:
\begin{equation}
\label{bosonic-CKP}
{\rm Res}_{z=0}\, 
\exp\left( \sum_{i=1}^\infty (t_i+(-1)^i {t'}_i)z^i\right)
\exp\left(- \sum_{i=1}^\infty \left(
\frac{\partial}{\partial  t_i}+(-1)^i\frac{\partial}{\partial  {t'}_i}\right)
\frac{z^{-i}}{i}\right)\tau (t)\tau (t')=0.
\end{equation}

A polynomial KdV tau-function is a KP tau-function that  is independent of the even times $t_{2j}$ for $j=1,2,\ldots$,  hence the automorphism $\iota_C$ fixes this tau-function and we have the following 
\begin{corollary}
A polynomial KdV tau-function is also a CKP tau-function.
\end{corollary}
This  corollary does not hold for the non-polynomial  KdV tau-functions,  which can depend exponentially on the even times;
    such a tau-function is not fixed by $\iota_C$.

  Let $a(z)=\sum_{i=-M}^N a_iz^{i-1}$ and 
$v^\pm= {\rm Res}_z a(z)\psi^{\pm}(\pm z)=\sum_{i=-M}^N (\pm 1)^{i}a_i\psi^{\pm}_{i-\frac12}$.  Since $\iota_C(v^\pm)=\pm v^\mp$, we have
$
\iota_C(v^+v^-)=-v^-v^+=-(v^-,v^+)+ v^+v^-=v^+v^-
$,
because $(v^-,v^+)=0$.
Using this observation, we are now ready to prove the main result of this section.
\begin{theorem}
\label{prop5}
(a) Let $v_i^+\in \Psi^+$ for $i=1,\ldots ,k$, then
\begin{equation}
\label{main}
\tau_{v_1^+,\ldots ,v_k^+}(t):=
\langle (\exp H(t))v^+_1\iota_C(v^+_1)v^+_2\iota_C(v^+_2)\cdots v^+_k\iota_C(v^+_k)\rangle
\end{equation}
is a CKP tau-function.  For  
\begin{equation}
\label{v_j}
v_j^+=
{\rm Res }_z z^{ -a_j-1} \exp(\sum_{i=1}^\infty  c_{i,j}z^i)\psi^+(z),\quad j=1,\ldots,k
,
\end{equation}
this tau-function is, up to a sign,  equal to
\begin{equation}
\label{tau_v}
\det 
\begin{pmatrix}
T_{i,j}(t)
\end{pmatrix}_{1\le i,j\le k}, \ 
\mbox{where }
T_{i,j}(t)=\begin{cases}
\chi_{(a_i|a_j)}(t+c_i;t+\iota_C(c_j))&
i\le j,\\
  \iota_C(\chi_{(a_j|a_i)}(  t +\iota_C(c_j);t +c_i))&  i> j,
  \end{cases}
\end{equation}
where 
$c_j=(c_{1j},c_{2j},c_{3j},\ldots, c_{a_j+a_1+1,j})$ and 
$\iota_C(c_j)=(c_{1j},-c_{2j},c_{3j},\ldots, (-1)^{a_1+a_j}c_{a_j+a_1+1,j}))$,
with 
 $c_{nj}\in\mathbb C$,  $j=1, \ldots k$,  $n=1,2,\ldots$.
Here $\chi_{(a |b)}(s;t)$ is given by \eqref{chi},    \\
(b)
Any polynomial CKP tau-function is of the form  \eqref{main} and hence,
up to a   constant factor,     equal to \eqref{tau_v}.\\
(c) Any polynomial CKP tau-function is,
up to a   constant factor,  equal   to
\begin{equation}
\label{formT}
	\tau_{(a_1,\ldots,  a_k|a_1,\ldots,  a_k)}(t)=
 \det \left(
 \chi_{(a_i|a_j)}(t+c_i;t+\iota_C(c_j))
\right)_{1\le i,j\le k},
\end{equation}
where   the constants $c_j=(c_{1j},c_{2j},c_{3j},\ldots, c_{a_j+a_1+1,j})$ for $1\le i<j\le k$, must satisfy the following constraints:
\begin{equation}
\label{restrict2}
c_{a_i+a_j+1,j}=(-1)^{a_i+a_j}s_{a_i+a_j+1}
(c_{1,i}-c_{1,j}, c_{2,i}+c_{2,j}, \ldots ,c_{a_i+a_j,i}+(-1)^{a_i+a_j} c_{a_i+a_j,j},c_{a_i+a_j+1,i}).
\end{equation}
\\
(d) Any polynomial CKP tau-function  satisfies the following equation:
\begin{equation}
\label{Schur-CKPtau}
\sum_{k=0}^\infty s_k(\pm 2t_e) s_{k+1}(\mp \tilde\partial_e)\tau(t)=0,
\end{equation}
were  $t_e=(t_2,t_4,t_6,\ldots)$ and 
$\tilde\partial_e= (  \frac{\partial}{\partial t_2},\frac12 \frac{\partial}{\partial t_4},\frac13 \frac{\partial}{\partial t_6},\ldots)$.  
\end{theorem}

{\bf Proof.}
(a) It is obvious that $\tau_{v_1^+,\ldots ,v_k^+}(t)$ is a CKP tau function.  Let us calculate its explicit form for the $v_j^+$ given 
by \eqref{v_j}.  If 
\[
 v_j^+
 ={\rm Res }_z b_j(z)\psi^+(z),\quad \mbox{then }\iota_C(v_j^+)={\rm Res }_z b_j(z)\psi^-(-z)
 =-{\rm Res }_z b_j(-z)\psi^-(z),
 \]
where we take 
 \[
 b_j(z)
 =z^{  -a_j-1} \exp(\sum_{i=1}^\infty  c_{i,j}z^i).
\]
Then
\[
\begin{aligned}
b_j(-z)&=(-z)^{  -a_j-1} \exp(-\sum_{i=1}^\infty  -(-1)^ic_{i,j}z^i)\\
&=
(-1)^{  a_j+1}z^{  -a_j-1} \exp(-\sum_{i=1}^\infty  \iota_C(c_{i,j})z^i).
\end{aligned}
\]
Since  ${\rm Res}_z b_j(z)\psi^\pm (z)= \psi^\pm_{-a_j-\frac12}+\sum_{i>-a_j} g_i\psi^{\pm}_i$  for some $g_i\in\mathbb C$(cf. Lemma \ref{lemma1}(e)),  we find that
 $\lambda=(a_1,\ldots a_k|a_1,\ldots a_k)$.
 Using Wick's theorem, in the second equality below,  we find that  $\tau_{v_1^+,\ldots v_k^+}(t)$,  with $v_j^+$ given by \eqref{v_j}, is equal,  up to a possible sign, to 
 \[
 \begin{aligned}
 \langle& (\exp H(t))v_1^+\iota_C(v_1^+)\cdots v_k+\iota_C(v_1^+)\rangle
 \\
 =&
 {\rm Res}_{y_1}{\rm Res}_{z_1}\cdots {\rm Res}_{y_k}{\rm Res}_{z_1}\langle \psi^+(y_1)\psi^-(z_1)\cdots \psi^+(y_k)\psi^-(z_k)\rangle \prod_{j=1}^k b_j(y_j)b_j(-z_j)\\
 =&
 {\rm Res}_{y_1}{\rm Res}_{z_1}\cdots {\rm Res}_{y_k}\prod_{j=1}^k b_j(y_j)b_j(-z_j)\times
 \\
 Pf&
 \begin{pmatrix}
 0&
  \frac{e^{\sum_n  t_n(y_{1}^n-z_{1}^n)}}{y_{1}-z_{1}}&
  0&
  \frac{e^{\sum_n  t_n(y_1^n-z_{2}^n)}}{y_1-z_{2}}&0&\cdots& 0&\frac{e^{\sum_n  t_n(y_1^n-z_{k}^n)}}{y_1-z_{k}}\\
*&0&\frac{e^{\sum_n  t_n(y_{2}^n-z_{1}^n)}}{z_{1}-y_{2}}&0&\frac{e^{\sum_n  t_n(y_3^n-z_{1}^n)}}{z_1-y_{3}}&\cdots &\frac{e^{\sum_n  t_n(y_k^n-z_{1}^n)}}{z_1-y_{k}}&0\\
  0&*&
  0&
  \frac{e^{\sum_n  t_n(y_2^n-z_{2}^n)}}{y_2-z_{2}}&0&\cdots& 0&\frac{e^{\sum_n  t_n(y_2^n-z_{k}^n)}}{y_2-z_{k}}\\
  *&0&*&0&
  \frac{e^{\sum_n  t_n(y_3^n-z_{2}^n)}}{z_2-y_{3}}&\cdots &\frac{e^{\sum_n  t_n(y_k^n-z_{2}^n)}}{z_2-y_{k}}&0\\
\vdots&\vdots&\vdots&\vdots&\vdots&&\vdots&\vdots\\
 *&0&*&0&*&\cdots&0&\frac{e^{\sum_n  t_n(y_k^n-z_{k}^n)}}{y_k-z_{k}}
 \\
 0&*&0&*&0&\cdots&*&0
 \end{pmatrix}
 ,
 \end{aligned}
 \]
 for $|y_1|>|z_1|>|y_2|>\cdots>|z_k|$.  
 Next permuting rows and columns in the Pfaffian,  this is equal to
 \begin{equation}
 \label{XXX}
 \pm {\rm Res}_{y_1}{\rm Res}_{z_1}\cdots {\rm Res}_{y_k}\prod_{j=1}^k b_j(y_j)b_j(-z_j)\det\left(
  \frac{e^{\sum_n  t_n(y_{i}^n-z_{j}^n)}}{y_{i}-z_{j}}\right)_{1\le i,j\le k}.
  \end{equation}
 Using that for $i\le j$,  thus $|y_i|>|z_j|$,    we have 
 \[
 \begin{aligned}
 {\rm Res}_{y_i}{\rm Res}_{z_j}&b_i(y_i)b_j(-z_j)\frac{e^{\sum_n  t_n(y_{i}^n-z_{j}^n)}}{y_{i}-z_{j}}
 =\\
  {\rm Res}_{y_i}{\rm Res}_{z_j}&
  y_i^{  -a_j-1} (-z_j)^{  -a_j-1}
   \exp\sum_{k=1}^\infty  c_{k,i}y_i^k
  \exp\sum_{\ell=1}^\infty  c_{\ell,j}(-z_j)^\ell
  \frac{e^{\sum_n  t_n(y_{i}^n-z_{j}^n)}}{y_{i}-z_{j}}=\\
  {\rm Res}_{y_i}{\rm Res}_{z_j}&
  y_i^{  -a_j-1} (-z_j)^{  -a_j-1}
   \frac{e^{\sum_n  (t_n+ c_{ni})y_{i}^n-(t_n+\iota_C(c_{nj}) z_{j}^n}}{y_{i}-z_{j}}=\\
 \\
 &=(-1)^{a_j+1} \chi_{(a_i|a_j)}(t+c_i;t+\iota_C(c_j)),
 \end{aligned}
 \]
 and that for $i> j$,  thus $|y_i|<|z_j|$, 
 \[
 {\rm Res}_{z_j}{\rm Res}_{y_i}b_i(y_i)b_j(-z_j)\frac{e^{\sum_n  t_n(y_{i}^n-z_{j}^n)}}{y_{i}-z_{j}}=\pm
  \iota_C(\chi_{(a_j|a_i)}(  t +\iota_C(c_j);t +c_i)),
  \]
 we find that  \eqref{XXX} is equal, up to some sign, to \eqref{tau_v}.
 
 (b) Any polynomial CKP tau-function corresponds to a Lagrangian subspace  $L\subset \Psi^+$ with respect to $\omega$; it is of the form  $$L= {\rm span} \{ w_1^+,w_2^+,\cdots ,w_k^+\}\oplus\bigoplus _{j>k}\mathbb C \psi^+_j.
 $$
 Since $0=\omega(w^+_i,w^+_j)=(w^+_i, \iota_C(w^+_j))$, we find that $w^+_i$ and $\iota_C(w^+_j)$ anticommute for all $1\le i,j\le k$.      Hence,
 the corresponding tau-function is then equal,  up to a constant factor, to 
$ \tau_{w_1^+,\ldots ,w_k^+}(t)$.  This proves part (b).

(c) Let $\tau(t)$ be a  polynomial CKP tau-function,  i.e.  $\tau(t)$ is a KP tau-function that satisfies   
 \eqref{tauC}.  This $\tau$ is invariant under $\iota_C$.  We know from Section \ref{S3}  that $\sigma^{-1}(\tau(t))=f=R(A)f_\lambda\in R(U )f_\lambda$,  for some partition $\lambda=(\lambda_1,\lambda_2, \ldots,\lambda_\ell)=
 (a_1,\ldots ,a_k|b_1,\ldots, b_k)\in Par_\ell$.
 Thus $f$ is  of the form (cf. \eqref{RR1})
 \begin{equation}
 \label{f1}
f=
w_{\lambda_1-\frac12}^+ w_{\lambda_2-\frac32}^+\cdots w_{\lambda_\ell-\ell+\frac{1}2}^+|-\ell\rangle,
\end{equation}
where  
\begin{equation}
\label{w_j^+bis}
\begin{aligned}
 w_{\lambda_j-j+\frac{1}2}^+&=
  \psi^+_{j- \frac{ 1}2-\lambda_j}+ \sum_{  j+\frac12 -\lambda_j \le i<\ell }  a_{-i,\lambda_j-j+\frac{1}2}\psi_i^+\\
\end{aligned}
\end{equation}
and 
\begin{equation}
\label{orderM}
a_1={\lambda_1-1}>  \cdots >
a_k ={\lambda_k-k }\ge 0>{\lambda_{k+1}-k-1}>\cdots>
{\lambda_\ell- \ell }.
\end{equation}
Recall that  
\begin{equation}
\label{58a}
\iota_C(w_{\lambda_j-j+\frac{1}2}^+)=(-1)^{\lambda_j-j}\psi^-_{ j-\frac{1}2-\lambda_j}+ \sum_{  j+\frac12 -\lambda_j \le i<\ell }  (-1)^{i+\frac12}a_{-i,\lambda_j-j+\frac{1}2}\psi_i^-.
\end{equation}

Now let us study ${\rm Ann}\, f$, which is a maximal isotropic subspace of $\Psi$.
From \eqref{f1}, \eqref{58a},  and the fact that $\iota_C({\rm Ann}_+ f)={\rm Ann}_- f$, we deduce that 
\[
\begin{aligned}
	{\rm Ann}_+ f&={\rm span} \{ w_{\lambda_1-\frac12}^+, w_{\lambda_2-\frac32}^+,\ldots,w_{\lambda_\ell-\ell+\frac12}^+,  \psi_{\ell+\frac12}^+, \psi_{\ell+\frac32}^+, \ldots\},
\\
{\rm Ann}_- f&={\rm span} \{ \iota_C(w_{\lambda_1-\frac12}^+,) \iota_C(w_{\lambda_2-\frac32}^+),\ldots,\iota_C(w_{\lambda_\ell-\ell+\frac12}^+),  \psi_{\ell+\frac12}^-, \psi_{\ell+\frac32}^-, \ldots\}.
\end{aligned}
\]
Thus  $(\psi_i^-, w_{\lambda_j-j+\frac12}^+)=0$ for all $i>\ell$, and  $j=1,\ldots ,\ell$,  which means that $a_{-i,\lambda_j}=0$ for all $i<-\ell$. 
Therefore  all $\lambda_j-j\le\ell$, for $j=1,\ldots,\ell $.  In particular $a_1\le \ell$.

Next, consider the element
\begin{equation}
\label{gg}
	g=w_{\lambda_1-\frac12}^+\iota_C(w_{\lambda_1-\frac12}^+)w_{\lambda_2-\frac32}^+\iota_C(w_{\lambda_2-\frac32}^+)\cdots w_{\lambda_k-k+\frac12}^+\iota_C(w_{\lambda_k-k+\frac12}^+)|0\rangle,
\end{equation}
where $k$ is determined by  the Frobenius notation of $\lambda$.
Note that this  element is not equal to $0$,  because the subspace  spanned by $w_{\lambda_1}^+,\iota_C(w_{\lambda_1}^+), \ldots, w_{\lambda_k-k+\frac12}^+,\iota_C(w_{\lambda_k-k+\frac12}^+)$ is isotropic with respect to $\omega(\cdot,\cdot)$ and   all elements  $w_{\lambda_j-j+\frac12}^+$ and $\iota_C(w_{\lambda_j-j+\frac12}^+)$ for $j=1,\ldots, k$  do not annihilate  $|0\rangle$.
First of all consider $\psi_i^\pm g$ for $i>\ell$.  Since $\psi_i^\pm\in {\rm Ann}_\pm f$,  this element anticommutes with all $w^+_{\lambda_j-j+\frac12}$ and $\iota_C(w^+_{\lambda_j-j+\frac12})$ for $j=1,\ldots, \ell$.  Thus 
\[
\psi_i^\pm g=w_{\lambda_1-\frac12}^+\iota_C(w_{\lambda_1-\frac12}^+)w_{\lambda_2-\frac32}^+\iota_C(w_{\lambda_2-\frac32}^+)\cdots w_{\lambda_k-k-\frac12}^+\iota_C(w_{\lambda_k-k-\frac12}^+)\psi_i^\pm|0\rangle=0.
\]
Also   $w^+_{\lambda_j-j+\frac12}$ (resp. $\iota_C(w^+_{\lambda_j-j+\frac12})$) anticommutes with  $\iota_c(w^+_{\lambda_j-i-\frac12})$   (resp.  $w^+_{\lambda_j-i+\frac12}$).
If   $j\le k$,  then 
 \[
 w_{\lambda_j-j+\frac12}^+g=w_{\lambda_1-\frac12}^+\iota_C(w_{\lambda_1-\frac12}^+)
 \cdots w_{\lambda_j-j+\frac12}^+w_{\lambda_j-j+\frac12}^+\iota_C(w_{\lambda_j-j+\frac12}^+)\cdots
 w_{\lambda_k-k+\frac12}^+\iota_C(w_{\lambda_k-k+\frac12}^+)|0\rangle=0,
 \]
 and similarly $\iota_C(w_{\lambda_j-j+\frac12}^+)g=0$ for $j\le k$.  
 Next,   let $j>k$,  then also
 \[
 w_{\lambda_j-j+\frac12}^+g=w_{\lambda_1-\frac12}^+\iota_C(w_{\lambda_1-\frac12}^+)w_{\lambda_2-\frac32}^+\iota_C(w_{\lambda_2-\frac32}^+)\cdots  
 w_{\lambda_k-k+\frac12}^+\iota_C(w_{\lambda_k-k+\frac12}^+)w_{\lambda_j-j+\frac12}^+|0\rangle=0,
 \]
 since $\lambda_j-j<0$ and $w_{\lambda_j-j+\frac12}^+= \psi^+_{j-\lambda_j-\frac12}+\sum_{ i>j-\lambda_j>0}a_{-i, \lambda_j-j+\frac12}\psi^+_i$,   i.e.  a linear combination of  $\psi^+_i$ with $i>0$.
 This also holds for $\iota_C(w^+_{\lambda_j-j+\frac12})$ for $j>k$.
 Hence ${\rm Ann}\, f={\rm Ann}\, g$ and $g$ must be a multiple of  $f$.
 Therefore,   
 $\tau(t)= {\rm const}\,  \tau_{w_{\lambda_1-\frac12}^+ w_{\lambda_2-\frac32}^+\cdots w_{\lambda_k-k+\frac12}^+}(t)$,  and  
 $b_j=a_j$ for all $1\le j\le k$.  Hence $\lambda$
 is self-conjugate and equal  to
 $( a_1,\ldots, a_k| a_1,\ldots, a_k)$,  in the Frobenius notation.
 
Note that in this construction,  the span of $w_{\lambda_1-\frac12}^+, \ldots, w_{\lambda_k-k+\frac12}^+$ is isotropic with respect to $\omega(\cdot,\cdot )$.  This gives some restriction on the $a_{i,\lambda_j-j+\frac12}$.  
 For arbitrary vectors $v_1^+,\ldots v^+_k$,  as in part (a) of the theorem,  the linear span of the $v_i^+$ does not have to be isotropic with respect to $\omega(\cdot,\cdot )$. 
 
 As in the previous  section,  we can write 
 \begin{equation}
 \label{W_j}
 \begin{aligned}
 w_{\lambda_j-j+\frac12}^+
 &=
 \psi^+_{ j-\frac12-\lambda_j}+ \sum_{  j+\frac12 -\lambda_j \le i<\ell }  a_{-i,\lambda_j-j+\frac{1}2}\psi_i^+\\
&={\rm Res }_z  z^{  j-\lambda_j-1} \psi^+(z)\exp(\sum_{i=1}^\infty  c_{i,j}z^i), \\
\iota_C(w_{\lambda_j-j+\frac12}^+)&={\rm Res }_z  
(-1)^{ \lambda_j-j}z^{  j-\lambda_j-1}\psi^-(z) \exp(-\sum_{i=1}^\infty  \iota_C(c_{i,j})z^i).
 \end{aligned}
 \end{equation}
 For these isotropic $w_{\lambda_j-j+\frac12}^+$ we have that
 \[
 \tau(t)=\langle (\exp H(t))w_{\lambda_1-\frac12}^+\cdots w_{\lambda_k-k+\frac12}^+ \iota_C(w_{\lambda_1-\frac12}^+)\cdots 
 \iota_C(w_{\lambda_k-k+\frac12}^+)\rangle
 \]
 So we can apply  Theorem \ref{T-Giambelli},  so that 
   $\tau(t)$ is given, up to a constant factor, by
 \eqref{tau-lambda2}, with $b_j=a_j=\lambda_j-j$ and  $d_j=\iota_C(c_j)$.  But there are restrictions on the constants, namely,
 since $\omega(w_{\lambda_i-i+\frac12}^+,w_{\lambda_j-j+\frac12}^+)=0$ for $1\le i<j\le k$,
 we find that 
 \[
 \begin{aligned}
 0&=\omega(w_{\lambda_i-i+\frac12}^+,w_{\lambda_j-j+\frac12}^+)\\
&={\rm Res }_y {\rm Res }_z   y^{ -a_i-1} z^{  -a_j-1}\exp(\sum_{n=1}^\infty ( c_{n,i}y^n -\iota_C(c_{n,j})z^n))
(\psi^+(y),\psi^-(z) )\\
&={\rm Res }_y {\rm Res }_z   y^{ -a_i-a_j-2}  \exp(\sum_{n=1}^\infty ( c_{n,i}  -\iota_C(c_{n,j}))y^n))
\delta(y-z)\\
&={\rm Res }_y   y^{ -a_i-a_j-2}  \exp(\sum_{n=1}^\infty ( c_{n,i}  -\iota_C(c_{n,j}))y^n))\\
&=s_{a_i+a_j+1}(c_i-\iota_C(c_j)).
 \end{aligned}
 \]
In other words, we obtain the restriction \eqref{restrict} for $b_j=a_j$ and $d_j=\iota_C(c_j)$, which means that we have to choose 
 $c_{a_i+a_j+1,j}$ for $j>i$ as in \eqref{restrict2}. 
\\
 (d) follows from   Remark \ref{R10} below.
 \hfill$\square$
 
\begin{remark}
\label{R10}
Following \cite{HHH}, \cite{AHH},  we can define $\iota_C$  on  $\Psi=\Psi^+\oplus \Psi^-$ as ${\rm ad}\, \Omega$, 
i.e.  $\iota_C(\psi^\pm_j)={\rm ad}\, \Omega (\psi^\pm_j)$
for
\[
\Omega=\frac12(\Omega_++\Omega_- ), \quad\mbox{where }
\Omega_\pm=\sum_{i\in\frac12+\mathbb Z} (-1)^{i\mp\frac12}\psi_i^\pm\psi^\pm_{-i} .
\]
Since $\Omega |0\rangle=0$,  we find that a CKP element $f\in F^{(0)}$ of the form 
\[
f=v^+_1\iota_C(v^+_1)v^+_2\iota_C(v^+_2)\cdots v^+_k\iota_C(v^+_k)|0\rangle
\]
satisfies $\Omega f=0$.
Since $\Omega_\pm f\in F^{(\pm 2)}$,  we must have that.
\begin{equation}
\label{Omegapm}
\Omega_\pm f=0.
\end{equation}
Now applying $\sigma$ to the above equations \eqref{Omegapm}, and using that
\[
\Omega_+= {\rm Res}_z \psi^+(-z)\psi^+(z), \quad \Omega_-= {\rm Res}_z \psi^-(z)\psi^-(-z),
\]
we find that   a polynomial  CKP tau function $\tau(t)$
satisfies
\[
{\rm Res}_z z 
\exp\left( \pm2\sum_{i=1}^\infty  t_{2i} z^{2i}\right)
\exp\left(\mp\sum_{i=1}^\infty 
\frac{\partial}{\partial  t_{2i} }
\frac{z^{-2i}}{i}\right)\tau (t) =0,
\]
which is,  in  terms of the elementary Schur polynomials,  equation \eqref{Schur-CKPtau}.
 \end{remark}
\begin{example}
To show that the constraints \eqref{restrict2} are non-trivial,  we calculate an example explicitly.The smallest example where this constraint occurs, is for $\lambda=(2,2)=(1,0|1,0)$. We have
$$
\begin{aligned}
\tau_{(1,0|1,0);c}(t)=&
1/12 (12 c_{21 }c_{22} - 12 c_{31} t_1 - 6 c_{21} t_1^2 + 6 c_{22} t_1^2 + t_1^4 - 
   2 c_{11}^3 (c_{12} + t_1) + \\
  &+ 6 c_{11} (c_{12}^2 t_1 + 
   2 (c_{22} - c_{21}) t_1 
     + c_{12} ( t_1^2 -2 c_{21 }- 2 t_2)) + 12 c_{21} t_2 + \\
     &+12 c_{22} t_2 + 12 t_2^2 + 
   3 c_{11}^2 (c_{12}^2 + 2 c_{22} - t_1^2 + 2 t_2) + 
   3 c_{12}^2 (2 c_{21} + t_1^2 + 2 t_2) \\
   &- 12 t_1 t_3 - 
   4 c_{12 }(3 c_{31} - t_1^3 + 3 t_3)).
    \end{aligned}
   $$
   Calculating $\tau_{(1,0|1,0);c}(t)-\tau_{(1,0|1,0);c}(\iota_C(t))$, we find that
   $$
   \tau_{(1,0|1,0);c}(t)-\tau_{(1,0|1,0);c}(\iota_C(t))=(c_{11}^2 - 2 c_{11} c_{12} + c_{12}^2 + 2 (c_{21} + c_{22})) t_2,
   $$
  and this term has to be 0 for a CKP tau-function.  This happens if  we  choose
  $$
 c_{22}=-( \frac12c_{11}^2 -  c_{11} c_{12} + \frac12c_{12}^2 +  c_{21} )
 =-s_2(c_{11}-c_{12},  c_{21})
  $$
  which is exactly the constraint \eqref{restrict2}.
\end{example}

Let $\tau(t)$ be as in 
\eqref{tau_v}.   We find  in a similar way as  we obtained \eqref{TTT},  that the corresponding CKP wave function is given by
\begin{equation}
\label{TTTT}
\begin{aligned}
&w^+(t,z)=\frac1{\tau(t)}\exp\left(\sum_{n=1}^\infty t_nz^n\right)\times \\
&\det
\begin{pmatrix}
1&\chi_{(0|a_1)}([z^{-1}];t+\iota_C(c_1))
&\cdots&\chi_{(0|a_k)}([z^{-1}];t+\iota_C(c_k))
\\
s_{a_1 }(t+c_1   )&T_{11}(t)
&\cdots&T_{1 k}(t)
\\
s_{a_2 }(t+c_2  )&T_{21}(t)
&\cdots&T_{2k}(t)
\\
\vdots&\vdots&
 &\vdots
\\
s_{a_k }(t+c_k)&T_{k1}(t)&
\cdots&T_{kk}(t)
\end{pmatrix}
,
\end{aligned}
\end{equation}
where the $T_{ij}(t)$ are  given by \eqref{tau_v}.

It follows from the bilinear identity \eqref{bosonic-mKP} on the tau-function, that 
the  wave function satisfies 
\begin{equation}
\label{82a}
{\rm Res}_z\, w^+(t,z)w^+(t',-z)=0.
\end{equation}
Writing   $w^+(t,z)$ as in \eqref{def P},  we obtain as in \eqref{82a}:
\[
\begin{aligned}
0=&
{\rm Res}_z\, P^+(t, z) P^+(t', -z)  \exp\left(\sum_{n=1}^\infty t_n z^n+{t'}_n(-z)^n\right) \\
=&{\rm Res}_z\, P^+(t, z) P^+(t' -z)  \exp\left(\sum_{n=1}^\infty (t_n -\iota_C({t'}_n))z^n\right),
\end{aligned}
\]
from which we deduce that $ P^+(t, \partial)^{-1}= P^+(\iota_C(t),\partial)^*$.
This 
implies   that, besides the Lax equations  \eqref{Laxeq}, the pseudodifferential operator 
 $L(t,\partial)=P^+(t, \partial)\circ\partial \circ P^+(t, \partial)^{-1}$ satisfies the condition 
\begin{equation}
\label{82b}
L(t,\partial)^*=(P^+(t, \partial)\circ\partial \circ P^+(\iota_C(t),\partial)^*)^*
=
-L(\iota_C(t), \partial).
\end{equation}
It follows from   \eqref{82b} that the pseudodifferential operator $L(t,\partial)$ is skew-adjoint if  
and only if 
$L(t,\partial)$ satisfies the Krichever-Zabrodin condition \cite{KZ} 
\[ 
  [ \frac{\partial}{\partial t_{2n}}, L(t,\partial)]\Big|_{t_{2n}=0}=0,\quad n=1,2,\ldots ;
\]
in particular, if
$L(t,\partial)=L(t_o,\partial)$, where $t_o=(t_1,0,t_3,0,t_5,\ldots)$. In the latter case the hierarchy \eqref{Laxeq} of Lax-Sato equations becomes
\begin{equation}
\label{86aa}
\frac{\partial L (t_o,\partial)}{\partial t_k}=[(L (t_o,\partial)^k)_+, L (t_o,\partial)],\quad k=1,3,5,,\ldots ,
\end{equation}

\section{Reductions of the CKP hierarchy}
\label{S6}
Let $n$ be a positive integer $\ge 2$. We want to study $n$-reductions of the CKP hierarchy,  which  means that we restrict $c_\infty$  to the Lie algebra   $c_\infty\cap \hat{gl}_n$,  where    $\hat{gl}_n= gl_n(\mathbb C[t,t^{-1}]\oplus K$
  is the subalgebra of $a_\infty$,  consisting of all $n$-periodic matrices  $g=(g_{ij})_{i,j\in\frac12+\mathbb{Z}}$,
    i.e. $g_{i+n,j+n}=g_{ij}$,
together with $K$. This intersection is equal to the affine Lie algebra  $\hat{sp}_{n}$ if $n$ is even, and to
the twisted affine Lie algebra $\hat{gl}_n^{(2)}$ if $n$ is odd (see \cite{JM},  page 977).

Let $G$ be an element in the  group ${\cal G}_n$, corresponding to this affine Lie algebra.
Then $G\psi_j^+G^{-1}= \sum_{j\in\frac12+\mathbb Z} a_{ij}\psi_i^+$,  where $(a_{ij})_{i,j\in\frac12+\mathbb Z}$ is periodic, i.e. 
$a_{i+n,j+n}=a_{ij}$,
and satisfies 
\[
(-1)^{j-\frac12}\delta_{j,-\ell}=\omega(\psi_j^+, \psi_\ell^+)=\sum_{i,k} a_{ij}a_{k\ell}\omega(\psi_i^+, \psi^+_k).
\]
Hence $\sum_{i\in\frac 12+\mathbb Z} (-1)^{i-\frac12}a_{ij}a_{-i, \ell}=(-1)^{j-\frac12}\delta_{j,-\ell}$ and thus also
$\sum_{j\in\frac 12+\mathbb Z} (-1)^{j+\frac12}a_{ij}a_{\ell,-j }=(-1)^{i+\frac12}\delta_{i,-\ell}$.
Now let $p$ be an arbitrary positive integer, then 
\[
\begin{aligned}
(G\otimes G)&\sum_{j\in\frac12+\mathbb Z}\psi^+_j\otimes (-1)^{j-pn-\frac12}\psi^+_{pn-j}
\\
&=
\sum_{j\in\frac12+\mathbb Z}(-1)^{j-pn-\frac12}(G\psi^+_jG^{-1})G \otimes (G\psi^+_{pn-j}G^{-1})G
\\
&=\sum_{i,j,k\in\frac12+\mathbb Z}(-1)^{j-pn-\frac12} a_{ij}a_{k, pn-j}\psi^+_i G\otimes  \psi_k^+G
\\
&=\sum_{i,j,k\in\frac12+\mathbb Z}(-1)^{j-pn-\frac12} a_{ij}a_{k-pn, -j}\psi^+_i G\otimes  \psi_k^+G
\\
&=(-)^{pn-1}
\sum_{i,k\in\frac12+\mathbb Z}(-1)^{i+\frac12}\delta_{i, pn-k}\psi^+_i G\otimes  \psi_k^+G
\\
&= 
\sum_{i,k\in\frac12+\mathbb Z}(-1)^{i-pn-\frac12} \psi^+_i G\otimes  \psi_{pn-i}^+G.
\end{aligned}
\]
Thus $G\otimes G$ commutes with the operator
\[
\sum_{j\in\frac12+\mathbb Z}\psi^+_j\otimes (-1)^{j-pn-\frac12}\psi^+_{pn-j}.
\]
Since 
\[
\sum_{j\in\frac12+\mathbb Z}\psi^+_j|0\rangle\otimes (-1)^{j-pn-\frac12}\psi^+_{pn-j}|0\rangle=0,
\]
we find that 
\[
\sum_{j\in\frac12+\mathbb Z}\psi^+_j G|0\rangle \otimes (-1)^{j-pn-\frac12}\psi^+_{pn-j}G|0\rangle=0.
\]
Stated differently, we find that
\[
{\rm Res}_z\, z^{pn}\psi^+(z)f\otimes  \psi^+(-z)f=0 \quad\mbox{for}\quad f\in {\cal G}_n|0\rangle,  \quad p=0,1,\ldots .
\]
The case $p=0$ is the CKP hierarchy.  Using the isomorphism $\sigma$, we obtain the $n$-reduced CKP hierarchy of bilinear equations on the tau-function ($p=0,1,2, \ldots$):
\begin{equation}
\label{bosonic-CKPred}
{\rm Res}_{z=0}\, 
z^{pn}\exp\left( \sum_{i=1}^\infty (t_i+(-1)^i {t'}_i)z^i\right)
\exp\left(- \sum_{i=1}^\infty \left(
\frac{\partial}{\partial  t_i}+(-1)^i\frac{\partial}{\partial  {t'}_i}\right)
\frac{z^{-i}}{i}\right)\tau (t)\tau (t')=0
.
\end{equation}
\begin{remark}
(a)
Since  $\tau(t)$ is also an $n$-reduced KP tau function we find that $\frac{\partial\tau(t)}{\partial t_{pn}}={\rm const} \tau(t)$, which gives for polynomial tau-functions that $\frac{\partial\tau(t)}{\partial t_{pn}}=0$.
\\
(b) Note that the above equations  \eqref{bosonic-CKPred}   on the tau-function induce  the following equations on the wave function $w^+(t,z)$:
\[
{\rm Res}_z\, z^{pn}w^+(t,z)w^+(t',-z)=0,\quad p=1,2,\ldots.
\]
Taking $p=1$, one deduces that a  reduced  Lax operator $\cal L$,  which we define as ${\cal L}(t,\partial)=L(t,\partial)^n$, is an $n$-th order monic  differential operator,
satisfying
\begin{equation}
\label{Laxeqred}
\frac{\partial {\cal L} (t,\partial)}{\partial t_k}=[({\cal L }(t,\partial)^{\frac{k}{n}})_+, {\cal L} (t,\partial)]
\quad\mbox{ and}\quad 
{\cal L}(t,\partial)^*=(-1)^n{\cal L}(\iota_C(t),\partial),
\end{equation}
where $k$ is a positive integer.  If $k$ is  divisible by $n$, the right-hand side of the first equation  is 0.
The second equation follows from \eqref{86aa}.
\end{remark}

We now want to construct all polynomial tau-functions of this $n$-reduced CKP hierarchy.  Recall the construction of Section \ref{S3},  and  let $G\in{\cal  G}_n$;  if we consider $G$  as an element of $SL_n(\mathbb C[t,t^{-1}])$, then,  using the results of \cite {KvdLmodKP},  the  element $G|0\rangle $ must lie in one of the 
Schubert cells $R(U)f_\lambda$ corresponding to a $\lambda\in Par_\ell$  that is {\it $n$-periodic},  which means that the sequence  
\[
V_\lambda:=\{\lambda_1-1,\lambda_2-2,\ldots ,\lambda_\ell-\ell ,-\ell-1,-\ell-2,\ldots\}
, 
\]
is mapped to itself if  we subtract $n$ from all its members.
For such an $n$-periodic partition $\lambda$ define the subset
\[
V^{(n)}_\lambda=\{   \lambda_1-1,\lambda_2-2,\ldots ,\lambda_\ell-\ell\}\backslash \{
\lambda_1-n-1,\lambda_2-n-2,\ldots ,\lambda_\ell-n-\ell\}.
\]
The cardinality of the set  $V_\lambda^{(n)}$ is at most $n$.  However if it is equal to $n$, then  $V_\lambda =\{ j\in\mathbb Z| \, j<\lambda_1\}$, and hence $\lambda=\emptyset $.  Hence,  without loss of generality we may assume that  the cardinality of 
$V_\lambda^{(n)}$ is equal to $m<n$.  If  $V_\lambda^{(n)}=\{ \mu_1>\mu_2>\cdots> \mu_m\}$, then 
\[
V_\lambda=\{\mu_i-sn|\, i=1,\ldots, m,\ s\in\mathbb Z_{\ge 0}\}.
\]
The fact that this KP tau-function should also be a CKP tau-function gives that $\lambda$ must be self-conjugate, thus
\[
\begin{aligned}
V_\lambda=&\{a_1,a_2,\ldots , a_k,-k-1,-k-2,\ldots, -a_k,  -a_k-2, \ldots,  -a_{k-1}, -a_{k-1}-2,\ldots\\
 &\qquad \ldots, -a_{2}, -a_{2}-2,\ldots , -a_1=1-\ell, -\ell-1,-\ell-2,-\ell-3\ldots \}
.
\end{aligned}
\]
{Both  restrictions, $n$-periodic and self-conjugate,  lead to  the following two conditions on the set $A_\lambda^{(n)}:=\{a_1,a_2,\ldots, a_k\}$:
\begin{itemize}
\item If $a_j\in A_\lambda^{(n)}$, then either $a_{j}-n\in A_\lambda^{(n)}$ or $a_j-n<0$.
\item For all $a_i, a_j\in A_\lambda^{(n)}$,   the integer $a_i+a_j+1$ is not a multiple of $n$.
\end{itemize}
}
Note that if $a_i+a_j+1=kn$,  for some positive integer $k$, then $a_i\in V^{(n)}_\lambda$ but $a_i-kn=-a_j-1\not\in  V^{(n)}_\lambda$.

We  prove this allong the lines of the proof of Theorem \ref{T-Giambelli}  and Theorem \ref{prop5},  assuming  that $\lambda\in Par_\ell$ is $n$-periodic and self-conjugate.
Then  $\lambda=(a_1,\ldots,a_k|a_1,\ldots,a_k)$ in the Frobenius notation, and 
\[
f_\lambda=\pm \prod_{i=1}^k\psi^+_{-a_i-\frac12}\iota_C(\psi^+_{-a_i-\frac12})|0\rangle   .
\]
Now let $G\in U$,  such that $G\psi_j^+G^{-1}= \sum_i a_{ij}\psi^+_i$,  where $A=(a_{ij})_{i,j\in\frac12+\mathbb Z}$, satisfying the condition that  $\sum_{i\in\frac 12+\mathbb Z} (-1)^{i-\frac12}a_{ij}a_{-i, \ell}=(-1)^{j-\frac12}\delta_{j,-\ell}$ (and thus also
$\sum_{j\in\frac 12+\mathbb Z} (-1)^{j+\frac12}a_{ij}a_{\ell,-j }=(-1)^{i+\frac12}\delta_{i,-\ell}$) and $a_{i+n,j+n}=a_{ij}$.
The first condition gives that the vectors 
\[
w^+_{-a_j-\frac12}=G\psi^+_{-a_j-\frac12}G^{-1}=\psi^+_{-a_j-\frac12}+\sum_{i>-a_j} a_{i,-a_j-\frac12}\psi_i
\] 
 for $j=1,\ldots,k$ form an isotropic subspace of $\Psi^+$ with respect to $\omega(\, ,\, )$.
Next, we investigate 
\[
G f_\lambda=w^+_{-a_1-\frac12}\cdots w^+_{-a_k-\frac12}\iota_C(w^+_{-a_1-\frac12})\cdots \iota_C(w^+_{-a_k-\frac12})|0\rangle.
\]
The $n$-periodicity of an element in $U$ gives  for $ a_j-n\ge 0$,    that  $a_j-n=a_r $ for an $r>j$, and  that 
$a_{i,-a_j-\frac12}$ is equal to  $a_ {i+n,-a_r-\frac12}.$  Hence
\[
\begin{aligned}
w^+_{-a_j -\frac12 }& =\psi^+_{-a_j-\frac12}+\sum_{i>-a_j } a_{i,-a_j-\frac12}\psi_i^+\\
& =\psi^+_{-a_j-\frac12}+\sum_{i>-a_j } a_{i+n,-a_r-\frac12}\psi_i^+\\
& =\psi^+_{-a_r+n-\frac12}+\sum_{i>-a_r } a_{i,-a_r-\frac12}\psi_{i+n}^+\\
\end{aligned}
\]
So if we assume, as in the proof of 
Theorem \ref{T-Giambelli},  that
\[
w^+_{-a_j-\frac12}={\rm Res}_z\, z^{-a_j-1}\psi^+(z) \exp\left(\sum_{i=1}^\infty c_{i,a_j} z^i\right),
\]
then
\[
w^+_{-a_r-\frac12}= w^+_{-a_j+n-\frac12}={\rm Res}_z\, z^{-a_j+n-1}\psi^+(z) \exp\left(\sum_{i=1}^\infty c_{i,a_j} z^i\right).
\]
This gives that there are at most $m$ different  vectors  $c_{a_j}=(c_{1,a_j}c_{2,a_j},\ldots)$.
So,  instead of $c_{a_j}$ we will write $c_{\overline {a_j}}$, where ${\overline {a_j}}$ stands for the congruence class of ${a_j}$ modulo $n$.  In a similar way as in Theorem \ref{prop5},  we obtain the restrictions on the constants.  This gives
\begin{theorem}
\label{T12}
Any polynomial $n$-reduced CKP tau-function is,
up to a   constant factor,  equal   to
\begin{equation}
\label{formTa}
	\tau_{(a_1,\ldots,  a_k|a_1,\ldots,  a_k);c}(t)=
 \det \left(
 \chi_{(a_i|a_j)}(t+c_{\overline{a_i}};t+\iota_C(c_{\overline{a_j}}))
\right)_{1\le i,j\le k},
\end{equation}
where  $\lambda=(a_1,\ldots,  a_k|a_1,\ldots,  a_k)$ is $n$-periodic and $c_{\overline{a_j}}=(c_{1,{\overline{a_j}}},c_{2,{\overline{a_j}}},c_{3,{\overline{a_j}}},\ldots, c_{m_j+\lambda_1,{\overline{a_j}}})\in\mathbb C^{m_j+a_1+1}$.  Here $m_j$ is the largest integer among all $a_1,\ldots, a_k$,
such that $\overline{m_j}=\overline{a_j}$. 
We have  the following restrictions  on the constants  for $1\le i<j\le k$:
\begin{equation}
\label{restrict2red}
s_{a_i+a_j+1}(c_{1,{\overline{a_i}}}-c_{1,{\overline{a_j}}}, c_{2,{\overline{a_i}}}+c_{2,{\overline{a_j}}}, \ldots  ,c_{a_i+a_j+1,{\overline{a_i}}}+(-1)^{a_i+a_j+1}c_{a_i+a_j+1,{\overline{a_j}}})=0.
\end{equation}
\end{theorem}

It is easy to see that the $n=2$-reduced CKP hierarchy \eqref{Laxeqred} coincides with the KdV hierarchy on the differential operator ${\cal L}=\partial^2+u$.

The next case, the $n=3$-reduced CKP hierarchy, is called the Kaup-Kupershmidt hierarchy. 
It is  the hierarchy \eqref{87a} of Lax equations on  the differential operator   
$\cal L$ given by
\eqref{87c}.
%
Since 
 $
({\cal L}^{\frac{5}{3}})_+=\partial^5+
\frac{5}{3} u\partial^3+\frac{5}{2} u_x\partial^2+
\frac{5}{18}
 (2 u^2 + 7 u_{xx}) \partial
 +
\frac{5}{9 }(uu_x+ u_{xxx})
$,
the first non-trivial  such equation occurs for $k=5$, and it gives
\begin{equation}
\label{87b}
\frac{\partial u}{\partial t_5}=-
\frac1{18}(
10 u^2 u_x  
 +25 u_xu_{xx} + 10 uu_{xxx}
  +2 u_{xxxxx}),
\end{equation}
which is the Kaup-Kupershmidt equation (see e.g. \cite{DKV}, Subsec.   11.3).

In this case there are, besides $\lambda=\emptyset$, two possible sets of self-conjugate partitions, which are $3$-periodic, viz. ($m\in\mathbb Z_{\ge0 }$):
\begin{equation}
\label{examplambda}
\begin{aligned}
(1)&\quad  \lambda=(3m,3m-3,3m-6,  \ldots,3,0|3m,3m-3,3m-6,  \ldots,3,0),\\
(2)&\quad \lambda=(3m+2,3m-1,3m-4,  \ldots,5,2|3m+2,3m-1,3m-4,  \ldots,5,2).
\end{aligned}
\end{equation} 
The corresponding CKP  tau-functions are equal,  up to  a constant factor, to,  respectively,
$$
\begin{aligned}
(1)&\quad 
 \det \left(
 \chi_{(3i|3j)}(t+c;t+\iota_C(c ))
\right)_{0\le i,j\le m},\\
 (2)&\quad  \det \left(
 \chi_{(3i+2|3j+2)}(t+c;t+\iota_C(c ))
\right)_{0\le i,j\le m},
\end{aligned}
$$
with  the following constraints on the vector of constants $c=(c_1,c_2,c_3,\ldots)$
$$
c_{2k}=-\frac12 s_{k}(2c_2,2c_4,2c_6, \ldots, 2c_{2k-4},2c_{2k-2},0),
$$
for $k=2, 5,8, \ldots,3m-4, 3m-1$,  and $k=4,7,10 , \ldots, 3m-2, 3m+1$, respectively.

Recall  that, by the second equation in   \eqref{Laxeqred}, ${\cal L}(t,\partial)^*=-{\cal L}(\iota_C(t),\partial)$.  Hence, in order to obtain a skew-adjoint differential operator,  one has to let all
$t_{2i}=0$, $i=1,2,3,\ldots$.
Also,  there is only one vector of constants, viz.  $c \in \mathbb C^{6m+1}$, and  $c= \in \mathbb C^{6m+5}$,  respectively.
 
 Note that due to the equation \eqref{25a},  which expresses  $u$ in terms of the tau-functions, the tau-functions  (1) and (2) with $t=t_o$  produce rational solutions of the Kaup-Kupershmidt hierarchy.

\section{Comparison with the polynomial solutions of BKP}
It is interesting to compare the polynomial tau-functions of the CKP hierarchy and that of the BKP
hierarchy.  The first observation is that both tau-functions  are parametrized more or less by the same kind of permutations.
The ones of the CKP are parametrized by the self-conjugate partitions $\lambda=(a_1,\ldots, a_k|a_1,\ldots, a_k)$,
where $a_1>a_2>\cdots>a_k\ge 0$.  Hence $\mu:=(a_1,a_2,\ldots, a_k)$ is an extended strict partition.  We use the word extended because we allow  $a_k$ to be zero.
The polynomial solutions of the BKP hierarchy are parametrized by the same set of extended strict partitions, with the only additional restriction   that $k$ has to be even.

The second observation  is that the solutions are expressed in terms of the polynomials: 
\begin{equation}
\label{chbar}
\begin{aligned}
\overline \chi_{M,N}(t,t')&=(-1)^N
\left( \frac 12 s_M(t)s_N(-t')+\sum_{k=1}^N  s_{M+k}(t)s_{N-k}(-t')\right)
\\
=&\chi_{(M,N)}(t,t')-(-1)^M\frac 12 s_M(t)s_N(-t').
\end{aligned}
\end{equation}
Namely, one has (see \cite {KvdLB2},  \cite {KRvdL} and  \cite{L})
\begin{theorem} \label{TheoB} 
(a) All polynomial tau-functions of the BKP hierarchy are,  up to a scalar
factor of the form
\begin{equation}
\label{TBKP}
\tau_{\lambda;c}^B( t_o)=Pf\left(\overline \chi_{\lambda_i\lambda_j}(t_o+c_i, t_o+\iota_C(c_j))\right)_{1\le i,j\le 2n},
\end{equation}
where $\lambda=(\lambda_1,\lambda_2,\ldots ,\lambda_{2n})$ is an extended strict partition, i.e. $\lambda_1>\lambda_2>\cdots \lambda_{2n}\ge 0$,  where $t_o=(t_1,0,t_3,0,t_5,0,\ldots)$ and $c_i=(c_{1i},c_{2i}, c_{3_i}, \ldots)$ are  arbitrary constants.\\
(b)  This tau-function is the square root of a KP tau-function $\tau_{\overline\lambda}( t_o)$,
where 
\[
\overline\lambda=
\begin{cases}
({\lambda_1-1},{\lambda_2-1},\cdots,{\lambda_{2n}-1}|{\lambda_1},{\lambda_2,}\cdots,{\lambda_{2n}}),&
\mbox{ if }\lambda_{2n}\ne 0\ \mbox{and}\\
({\lambda_1-1},{\lambda_2-1},\cdots,{\lambda_{2n-1}-1}|{\lambda_1},{\lambda_2,}\cdots,{\lambda_{2n-1}}),
&
\mbox{ if }\lambda_{2n}= 0.
\end{cases}
\]
\end{theorem}
Note the following:
\begin{itemize}
\item  In the above BKP tau-function,  the even times do not appear, but the even constants $c_{2k,i}$ do  appear.  
\item The formulas look different from the ones for the BKP tau-function in e.g.  \cite {KvdLB2}, but here we have used the fact that
\[
\begin{aligned}
\sum_{j=0}^\infty (-1)^j s_j(t)z^j&=\exp(\sum_{i=1}^\infty t_i(-z)^i)=\exp(-\sum_{i=1}^\infty \iota_C(t)z^i)=\sum_{j=0}^\infty  s_j(-\iota_C(t))z^j\\
\end{aligned}
\]
\item The square of $\tau_{\lambda}^B( t_o)$ is equal to
\[
\tau_{\lambda:c}^B( t_o)^2=\pm \det\left(\overline \chi_{\lambda_i\lambda_j}(t_o+c_i, t_o+\iota_C(c_j))\right)_{1\le i,j\le 2n}.
\] 
But since $\overline \lambda$ is not self-conjugate,  this is never equal to a CKP tau-function where one puts the even times equal to 0, except when $\lambda=\emptyset$, in that case $\tau_{0\emptyset ;c}^B( t_o)=$constant.
\end{itemize}

\end{document}